\documentclass[prc,reprint,twocolumn,showpacs,superscriptaddress]{revtex4-1}

\usepackage{graphicx}% needed for figures Include figure files
\usepackage{float}
\usepackage{dcolumn}   % needed for some tables
\usepackage{bm}        % for math
\usepackage{amssymb}   % for math add hypertext capabilities
\usepackage{mathtools}
\usepackage[english]{babel}
\usepackage{flafter}
\usepackage{footmisc}

\usepackage{hyperref}
\hypersetup{
colorlinks=true,
urlcolor= blue,
citecolor=blue,
linkcolor= blue%,
}

\makeatletter
\def\ignorecitefornumbering#1{%
     \begingroup
         \@fileswfalse
         #1%                     % do \cite comand
    \endgroup
}
\makeatother
\begin{document}
\title{The Structure of $^{33}$Si and the magicity of the N=20 gap at Z=14}

\author{S. Jongile}
%\email{sandilej@tlabs.ac.za}
\affiliation{Stellenbosch University, Private Bag X1, Matieland, Stellenbosch 7602 South Africa}
\affiliation{iThemba LABS, PO Box 722, Somerset West 7129, South Africa}

\author{A.~ Lemasson}
\affiliation{Grand Acc\'el\'erateur National d'Ions Lourds (GANIL),
CEA/DRF - CNRS/IN2P3, B.\ P.\ 55027, F-14076 Caen Cedex 5, France}
\affiliation{Department of Physics and Astronomy and National
Superconducting Cyclotron Laboratory, Michigan State University,
East Lansing, Michigan, 48824-1321, USA}

\author{O.~Sorlin}
\affiliation{Grand Acc\'el\'erateur National d'Ions Lourds (GANIL),
CEA/DRF - CNRS/IN2P3, B.\ P.\ 55027, F-14076 Caen Cedex 5, France}

\author{M.~Wiedeking}
\affiliation{iThemba LABS, PO Box 722, Somerset West 7129, South Africa}
\affiliation{School of Physics, University of the Witwatersrand, Johannesburg 2050, South Africa}

\author{P.~Papka}
\affiliation{Stellenbosch University, Private Bag X1, Matieland, Stellenbosch 7602 South Africa}
\affiliation{iThemba LABS, PO Box 722, Somerset West 7129, South Africa}

\author{D.~Bazin}
\affiliation{Department of Physics and Astronomy and National
Superconducting Cyclotron Laboratory, Michigan State University,
East Lansing, Michigan, 48824-1321, USA}

\author{C.~Borcea}
\affiliation{IFIN-HH, P. O. Box MG-6, 76900 Bucharest-Magurele, Romania}

\author{R.~Borcea}
\affiliation{IFIN-HH, P. O. Box MG-6, 76900 Bucharest-Magurele, Romania}

\author{A.~Gade}
\affiliation{Department of Physics and Astronomy and National
Superconducting Cyclotron Laboratory, Michigan State University,
East Lansing, Michigan, 48824-1321, USA}

\author{H.~Iwasaki}
\affiliation{Department of Physics and Astronomy and National
Superconducting Cyclotron Laboratory, Michigan State University,
East Lansing, Michigan, 48824-1321, USA}

\author{E.~Khan}
\affiliation{Institut de Physique Nucl\'eaire, IN2P3-CNRS,
F-91406 Orsay Cedex, France}
\affiliation{Universit\'e Paris-Saclay, Espace Technologique Bat. Discovery -
RD 128 - 2e \'et, 91190 Saint-Aubin, France}

\author{A.~Lepailleur}
\affiliation{Grand Acc\'el\'erateur National d'Ions Lourds (GANIL),
CEA/DRF - CNRS/IN2P3, B.\ P.\ 55027, F-14076 Caen Cedex 5, France}

\author{A.~Mutschler}
\affiliation{Institut de Physique Nucl\'eaire, IN2P3-CNRS,
F-91406 Orsay Cedex, France}
\affiliation{Grand Acc\'el\'erateur National d'Ions Lourds (GANIL),
CEA/DRF - CNRS/IN2P3, B.\ P.\ 55027, F-14076 Caen Cedex 5, France}

\author{F.~Nowacki}
\affiliation{Universit\'e de Strasbourg, CNRS, IPHC UMR 7178, F-67000 Strasbourg, France}

\author{F.~Recchia}
\affiliation{Department of Physics and Astronomy and National
Superconducting Cyclotron Laboratory, Michigan State University,
East Lansing, Michigan, 48824-1321, USA}

\author{T.~Roger}
\affiliation{Grand Acc\'el\'erateur National d'Ions Lourds (GANIL),
CEA/DRF - CNRS/IN2P3, B.\ P.\ 55027, F-14076 Caen Cedex 5, France}

\author{F.~Rotaru}
\affiliation {IFIN-HH, P. O. Box MG-6, 76900 Bucharest-Magurele, Romania}

\author{M.~Stanoiu}
\affiliation {IFIN-HH, P. O. Box MG-6, 76900 Bucharest-Magurele, Romania}

\author{S.~R.~Stroberg}
\affiliation{Department of Physics and Astronomy and National
Superconducting Cyclotron Laboratory, Michigan State University,
East Lansing, Michigan, 48824-1321, USA}
\affiliation{TRIUMF, 4004 Westbrook Mall, Vancouver, British Columbia, V67 2A3 Canada}

\author{J.~A.~Tostevin}
\affiliation{Department of Physics, University of Surrey,
Guildford, Surrey GU2 7XH, United Kingdom}

\author{M.~Vandebrouck}
\affiliation{Institut de Physique Nucl\'eaire, IN2P3-CNRS,
F-91406 Orsay Cedex, France}
\affiliation{Grand Acc\'el\'erateur National d'Ions Lourds (GANIL),
CEA/DRF - CNRS/IN2P3, B.\ P.\ 55027, F-14076 Caen Cedex 5, France}

\author{D.~Weisshaar}
\affiliation{Department of Physics and Astronomy and National
Superconducting Cyclotron Laboratory, Michigan State University,
East Lansing, Michigan, 48824-1321, USA}

\author{K.~Wimmer}
\affiliation{Department of Physics, The University of Tokyo, Hongo, Bunkyo-ku, Tokyo 113-0033, Japan}
\affiliation{Department of Physics, Central Michigan University, Mt. Pleasant, Michigan 48859, USA}
\affiliation{Department of Physics and Astronomy and National
Superconducting Cyclotron Laboratory, Michigan State University,
East Lansing, Michigan, 48824-1321, USA}

\begin{abstract}
The structure of $^{33}$Si was studied by a one-neutron knockout reaction from a $^{34}$Si beam at 98.5 MeV/u incident on a $^{9}$Be target. The prompt $\gamma$-rays following the de-excitation of $^{33}$Si were detected using the GRETINA $\gamma$-ray tracking array while the reaction residues were identified on an event-by-event basis in the focal plane of the S800 spectrometer at NSCL (National Superconducting Cyclotron Laboratory). The presently derived spectroscopic factor values, $C^2S$, for the 3/2$^+$ and 1/2$^+$ states, corresponding to a neutron removal from the $0d_{3/2}$  and
$1s_{1/2}$ orbitals, agree with shell model calculations and point to a strong $N=20$ shell closure. Three states arising from the more bound $0d_{5/2}$ orbital are proposed, one of which is unbound by about 930 keV. The sensitivity of this experiment has also confirmed a weak population of 9/2$^-$ and 11/2$_{1,2}^-$ final states, which originate from a higher-order process. This mechanism may also have populated, to some fraction, the 3/2$^-$ and 7/2$^-$ negative-parity states, which hinders a determination of the $C^2S$ values for knockout from the normally unoccupied $1p_{3/2}$  and $0f_{7/2}$ orbits.
\end{abstract}

\maketitle

\section{\label{sec:level1}INTRODUCTION}

The nucleus $^{34}$Si is fascinating in many aspects. It is proposed to exhibit a central proton density depletion, typically referred to as a ``bubble'' \cite{Grass09,Muts16,Dugu17}, and is also one of a small number of nuclei that experiences a drastic reduction of its spin-orbit splitting (here the $1p_{3/2} -1p_{1/2}$ splitting) \cite{Burg14} in comparison to the neighboring isotones.  It has the properties of a spherical, doubly-magic nucleus, e.g. a high 2$^+_1$ energy at 3325 keV and the tentative spherical 2$^+$ state at an even higher energy of 5348~keV \cite{Zege10,Lica19}, a low $B(E2; 0^+_1 \rightarrow 2^+_1)$ value \cite{Ibbo98}, and a drop in the neutron separation energy ($S_n$) beyond $N=20$ by about 5~MeV \cite{Asch19}. Its $N=20$ gap was corroborated by the energy of the $4^-,5^-$ states \cite{Lica19}.

In addition, $^{34}$Si lies at the shore of the island of inversion, where nuclei become deformed. For example, the nearby $^{34}$Al nucleus has ground and intruder configurations separated by only 46.6~keV \cite{Lica17}. It follows that deformed configurations, shape coexistence \cite{Rota12, Lica19}, and possibly triaxial shapes were predicted \cite{Han17} and searched for \cite{Lica19,Pasch11,Wied08} in the properties of the first few excited states of $^{34}$Si. The abrupt transition from the closed-shell ground state of $^{34}$Si to the intruder-dominated ground state of $^{32}$Mg \cite{Wimm10,Craw16,Terry08}, with only two protons removed from the $0d_{5/2}$ orbit, is attributed to the subtle balance between the magnitude of the proton and neutron shell gaps that prevent nuclear excitations and the pairing and quadrupole correlations that scale with the amount of particle-hole excitations across these gaps.

In order to investigate the extent of such correlations, single-nucleon transfer reactions in direct kinematics have been used for the past decades. These direct reactions have allowed one to extract spectroscopic factors (and hence occupation numbers in the sum-rule limit) of orbitals below and above the Fermi surface from measured partial transfer cross sections analyzed with direct reaction theory. When there is an abrupt drop in the occupancy of single-particle states above the Fermi surface, this Fermi surface is said to be ``stiff'', meaning that closed-shell effects are large enough to suppress the effects of correlations. The $^{40}$Ca$(p,d)^{39}$Ca reaction was used in Ref. \cite{Mato93} to probe the Fermi surface of $^{40}$Ca. The analysis concluded an almost fully occupied $0d_{3/2}$ orbital, with little occupancy of the $0f_{7/2}$ (0.14) and  $1p_{3/2}$ (0.01) orbits. Moreover, the energy centroids of the $1s_{1/2}$ and $0d_{5/2}$ hole states in $^{39}$Ca were determined to be 2.65 and 6.61 MeV, respectively, upon integrating the identified fragmented $1/2^+$ and $5/2^+$ strength up to 9.2 MeV.

One of the goals of the present work is to investigate the stiffness of the Fermi surface in the unstable $^{34}$Si nucleus by means of the $^9$Be($^{34}$Si,$^{33}$Si+$\gamma$)X one-neutron knockout reaction, where $\gamma$-ray detection is used to tag the final state. The energy of the hole states arising from the $1s_{1/2}$ and $0d_{5/2}$ orbitals will also be characterized. Prior to the present work, the knockout residue $^{33}$Si was studied using a variety of experimental techniques, e.g.  multi-nucleon transfer \cite{Fifi85}, $\beta$-decay \cite{Morton}, thick-target deep-inelastic reactions \cite{Wang10}, and a one-neutron knockout reaction \cite{Ende12}. The work of Fifield {\it et al.} \cite{Fifi85} led to the identification of excited states at 1.47, 2.0, 3.19, 4.13 and 5.48 MeV with an uncertainty of about 40 keV, but without spin assignments. The $\beta$-decay experiment of Morton {\it et al.} \cite{Morton} identified a $(3/2^+,5/2^+)$ state at 4341 keV. In addition, the work of Wang et al. \cite{Wang10} proposed one $(9/2^-)$ state at 3159 keV and two $(11/2^-)$ states at 4090 and 4931~keV. In the previous one-neutron removal experiment of Enders {\it et al.} \cite{Ende12}, the $\gamma$-ray energy resolution was limited by the use of a scintillator array and only two excited states could be identified and characterized: the excited $1/2^+$ (1.01 MeV) and $(5/2)^+$  (4.32 MeV) states on top of the $3/2^+$ ground state \cite{Ende12}. No indications of excitations across the $N=20$ gap, such as population of the excited negative-parity $7/2^-$ state at 1435 keV or the $(3/2)^-$ state at 1981 keV, were observed. This led the authors to conclude that the $N=20$ magicity is well preserved in $^{34}$Si.

In the present one-nucleon removal study, we overcome some of the limitations encountered in Ref.~\cite{Ende12} by using the high-energy resolution $\gamma$-ray tracking array GRETINA \cite{Weis17,Pasc13}. As in Ref.~\cite{Ende12}, the exclusive parallel momentum distributions of the $^{33}$Si reaction residues, obtained by gating on $\gamma$-ray transitions, were used to infer the orbital angular momentum ($\ell$ value) of the removed neutron populating a specific final state. Despite the short running period, of only two hours, the direct population of the $3/2^-$ negative-parity valence state, resulting from the partial occupancy of the $1p_{3/2}$ orbital, as well as two new $(5/2^+)$ states, corresponding to the neutron removal from the $0d_{5/2}$ orbital, were observed. In addition, the high-$j$ $9/2^-$ and $11/2^-_{1,2}$ states, first identified in Ref.~\cite{Wang10}, were weakly populated in the present experiment, most probably by a two-step reaction mechanism.

\begin{figure}
\includegraphics[width=\columnwidth]{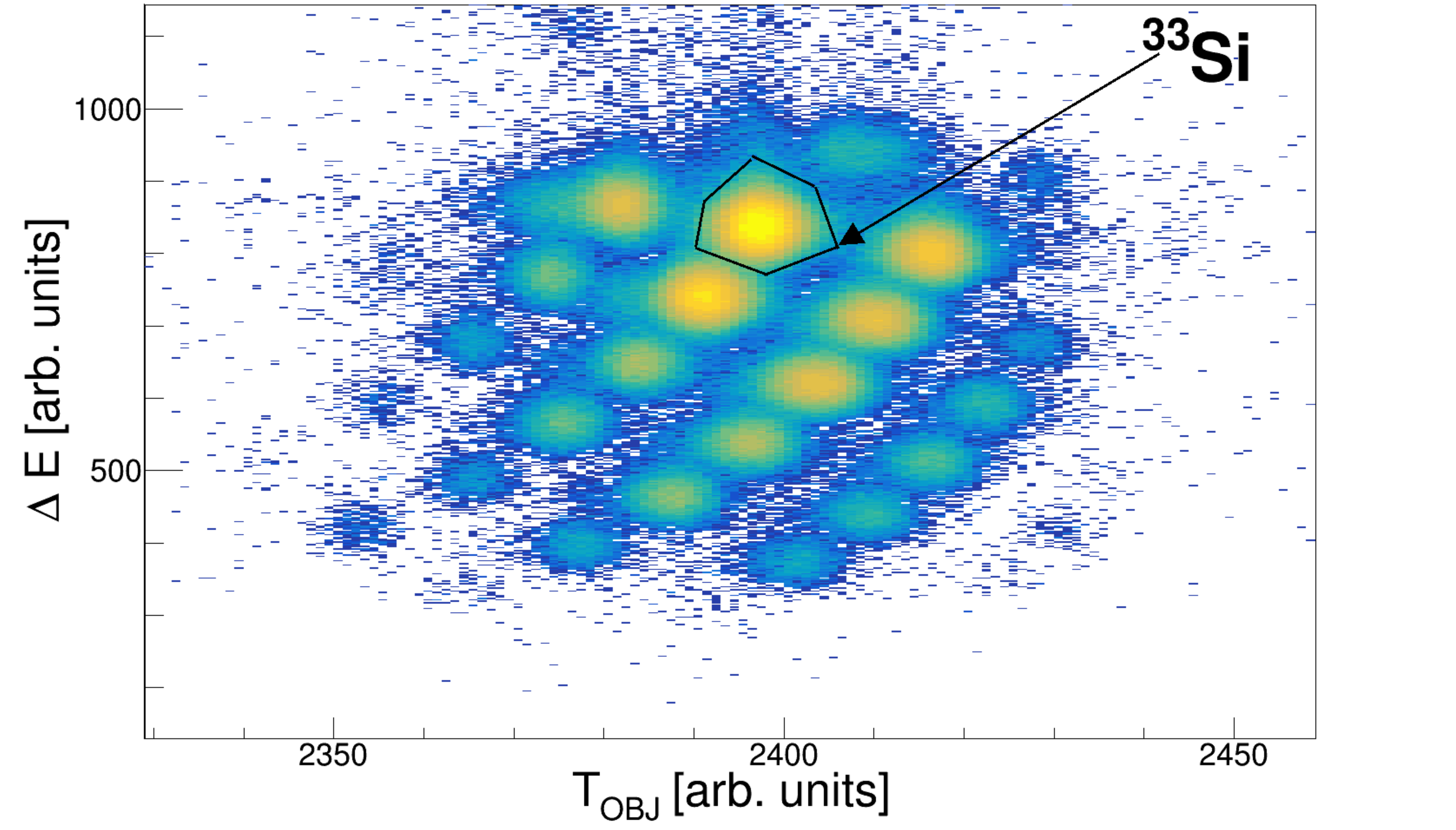}
\caption{(Color online) Identification of fragments using energy loss and
time-of-flight measurements using the S800 focal-plane detectors and plastic
timing detectors upstream of the reaction target. The cleanly identified
reaction residue of interest ($^{33}$Si) is indicated.}
\label{fig:epstart:tof}
\end{figure}

\section{\label{sec:level2}EXPERIMENTAL PROCEDURE}
The rare-isotope beam of $^{34}$Si was produced at the National Superconducting Cyclotron Laboratory (NSCL) from the fragmentation of a stable $^{48}$Ca beam at 140 MeV/u on a 846 mg/cm$^2$ thick $^{9}$Be target. The $^{34}$Si projectiles, produced at an average intensity of 4$\times10^5$ s$^{-1}$ and a purity of 70\%, were separated from all fragmentation products using the A1900 fragment separator \cite{Stolz,Morrisey} with an Al wedge of 300~mg/cm$^2$ thickness. The time-of-flight difference between plastic scintillators located at the extended focal  plane  of the A1900 and the object position of the S800 spectrograph \cite{Bazin2} was used for beam identification. The $^{34}$Si beam, at 98.5 MeV/u, then impinged on a $^9$Be reaction target (100~mg/cm$^2$) to produce the $^{33}$Si one-neutron knockout
residues at a velocity of $v = 0.4c$. The reaction residues emerging from the target were identified event-by-event with the S800 spectrograph by means of their energy loss measured in  the S800 ionization chamber located in the focal plane of the spectrograph and their time-of-flight taken between two scintillators placed at the object position and the focal plane of the S800 spectrograph. A typical identification matrix is shown in Fig.~\ref{fig:epstart:tof}. The two cathode readout drift chambers (CRDCs) \cite{Bazin} of the focal-plane detection system were used to reconstruct the trajectories of the reaction products, including $^{33}$Si, from position and angle measurements. The ion optics code (COSY) \cite{COSY} was used to generate an inverse map which allowed reconstruction of the non-dispersive position and momentum vectors at the S800 target position from the measured focal-plane parameters.

Prompt $\gamma$ rays emitted by $^{33}$Si were detected with the GRETINA (Gamma-Ray Tracking In-Beam Nuclear Array) array~\cite{Weis17,Pasc13} surrounding the $^9$Be reaction target. GRETINA was composed of seven detector modules, each consisting of four high-purity Ge crystals with 36 segments each. Four GRETINA modules were placed at $\theta_{lab}$ = $58^{\circ}$ and three at $\theta_{lab}$ = $90^{\circ}$. The $\gamma$-ray detection angles, derived from the $\gamma$ interaction positions obtained via the signal decomposition algorithm of GRETINA, were combined with the velocity vectors of the $^{33}$Si nuclei to perform event-by-event Doppler reconstruction of the $\gamma$-ray emitted in flight. For this,
the $\gamma$-ray interaction point with the highest energy deposition was considered to be the first hit entering the angle determination for the Doppler reconstruction. An in-beam energy resolution of $\sigma\simeq 2$ keV at 1~MeV was obtained. An add-back procedure was performed by summing the energies of several $\gamma$-ray interactions detected within neighboring crystals. An absolute in-flight efficiency of 6.5\% at 1 MeV was obtained from GEANT4 simulations \cite{Agos03} to account for the Lorentz boost. The simulation was benchmarked against standard calibration sources. 

The error on the centroid value originates from many factors among which are the uncertainties on the reconstructed impact point on target, secondary beam velocity and detection angle in the GRETINA array. The quoted uncertainty in Table \ref{tab:tabexp}, includes both statistical and systematic errors.  The experiment that is presented here is part of an experimental campaign in which several nuclei were studied in the same experimental conditions (see e.g. \cite{Muts16,Muts16b}).  A systematic uncertainty on the $\gamma$-ray energy centroid is estimated to be less than 2~keV at 4~MeV based on the comparison made in prior work using the same dataset and analysis procedure \cite{Muts16,Muts16b}. Moreover, we have studied the $^{36}$S(-1n)$^{35}$S experiment as well up to the neutron emission threshold, providing a broad range of $\gamma$-ray energies. The error bar is based on the comparison between the energies of already known states (with less than 1 keV error bar) to energies observed in our work, after having applied the Doppler correction.

\begin{figure*}
\includegraphics[width=\linewidth]{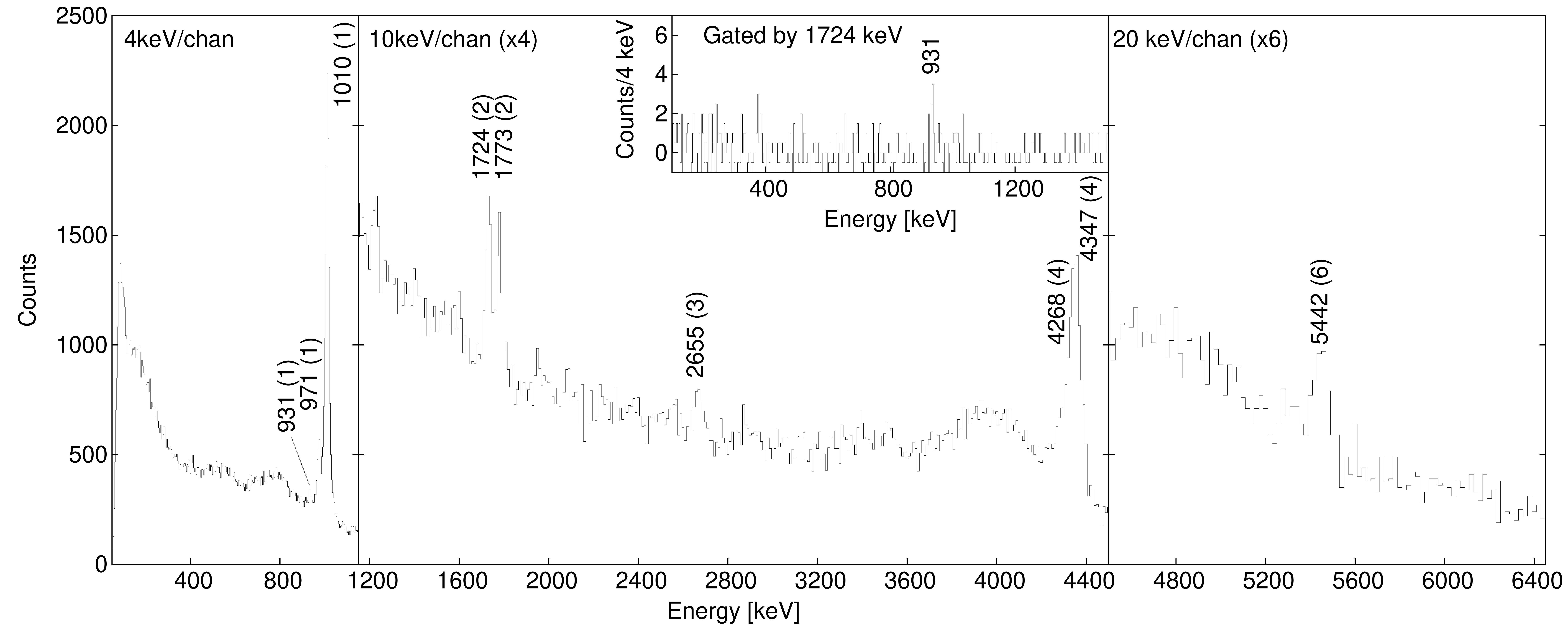}
\caption{(Color online) Doppler-reconstructed added-back spectra showing $\gamma$ rays detected using GRETINA in coincidence with $^{33}$Si identified with the S800 focal-plane detectors. The inset shows the $\gamma$-ray spectra in coincidence with the 1724~keV transition.}
\label{Egamma}
\end{figure*}

\begin{table*}
\caption{\label{tab:tabexp} Energy, spin and parity, population fraction $b_f$, and partial cross sections of
final states populated in the $^{34}$Si(-1n)$^{33}$Si reaction. Spectroscopic factors deduced from the present
work those quoted in Ref.\  \cite{Ende12} are also shown.}
\begin{ruledtabular}
\begin{tabular}{ccccccc}
Energy & $J^{\pi}$ & $b_f$\footnotemark[1]&  $\sigma^{inc}_{exp}\times$ $b_f$ & $\sigma_{sp}$ &  $C{^2}S_{exp} $ &  $C{^2}S$ Ref.~\cite{Ende12} \\
$[keV]$&  & $[\%]$ && [mb]&  & \\
\hline
0      & $\frac{3}{2}^+$   & 50.8 (55)  &  58.9 (71) &  15.1 & 3.90 (47) & 4.5 (7)  \\
1010(1)& $\frac{1}{2}^+$   & 22.8 (7)  &  26.4 (16)  &  19.7 & 1.34 (8) & 2.0 (3)  \\
1435(1)& $\frac{7}{2}^-$   & 9.0 (54)   &  10.4 (62) &  14.4 & 0.72 (43)\footnotemark[2] &  \\
1981(1)& $\frac{3}{2}^-$   & 4.5 (6)   &  5.2 (7)   &  14.6 & 0.35 (5)\footnotemark[2] &  \\
4268(4)& $(\frac{5}{2}^+)$ & 1.7 (3)   &  2.0 (3)   &  13.3 & 0.15 (3) &            \\
4347(4)& $(\frac{5}{2}^+)$ & 7.6 (5)   &  8.8 (6)   &  13.3 & 0.66 (6) & 1.3 (4)  \\
5442(6)& $(\frac{5}{2}^+)$ & 1.1 (3)   &  1.2 (4)   &  12.6 & 0.10 (3)\footnotemark[3]&   \\
\end{tabular}
\end{ruledtabular}
\footnotetext[1]{Values all normalized to 100\% of the one-neutron removal reaction.}
\footnotetext[2]{Upper limit assuming that these states are only populated by one-neutron knockout. See discussion in Sect.~\ref{sec:level6}}
\footnotetext[3]{Lower limit based on observed $\gamma$ transition from state above $S_n$.}
\end{table*}

\section{\label{sec:level22}Level Scheme}

\begin{figure}
\includegraphics [width=\columnwidth] {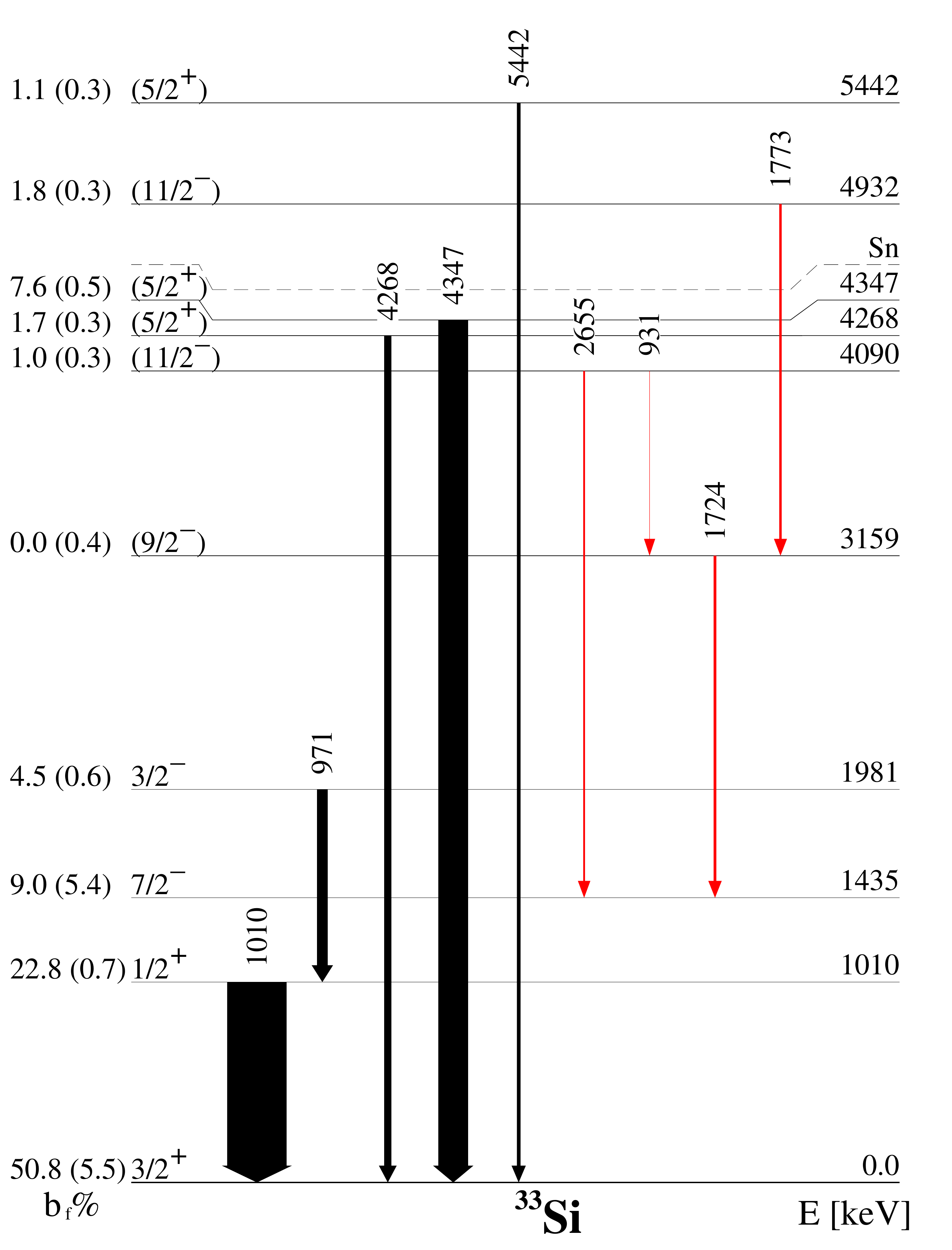}
\caption{(Color online) Experimental level scheme obtained in this work. }
\label{LS}
\end{figure}

The Doppler-corrected $\gamma$-ray spectrum in coincidence with $^{33}$Si is shown in Fig.~\ref{Egamma}. The corresponding level scheme, displayed in Fig.~\ref{LS}, was constructed using previously known information and, when available at sufficient statistics, $\gamma\gamma$ coincidence relationships deduced from a $\gamma\gamma$ coincidence matrix using add-back. We adopt the centroid energy values from the literature when reported with a better precision than the present measurement \cite{Ensdf}. When an observed $\gamma$-ray transition is close to or above the neutron separation energy of
$S_{n}$=4.5 MeV, it is tentatively proposed to decay directly to the $3/2^+$ ground state. In this level scheme, the transitions marked in black are expected to originate from the removal of a neutron from the occupied $sd$- or the $pf$-shell valence orbits. Removal from these orbits can directly populate states up to $j$=$5/2^+$ or $7/2^-$, respectively. Transitions shown in red connect higher-spin states, obviously requiring another reaction mechanism that will be discussed later. Most of the presently-identified $\gamma$ rays were perviously known, except for those discussed in the following.

As shown in the inset of  Fig \ref{Egamma}, the 931 keV transition is observed in coincidence with the 1724 keV transition, which establishes that the $(11/2^-_1)$ state at 4090 keV decays both to the $(9/2^-)$ state at 3159 keV and to the previously observed 10.2 ns isomeric $(7/2^-)$ state with an energy of 2655 keV~\cite{Wang10}. This new information on the competing decay branches strengthens the earlier, tentative assignment of $(11/2^-_1)$ for the state~\cite{Wang10}. A doublet of states is observed at 4268 and 4347 keV, of which the 4268 keV transition is observed for the first time. The 5442 keV transitions may correspond to the state at 5480 keV identified in Ref.~\cite{Fifi85}.
Interestingly, we observe the $\gamma$ decay of two states above $S_{n}$ (at 4932 and 5442 keV), states for which the open neutron-decay channel is to the $^{32}$Si ground state. Thus, these states are likely to have structures, or sufficiently large $j$ values (and centrifugal barriers) that their neutron decay is hindered. The former is true for the core-coupled $(11/2^-_2)$ state, expected to contain a negligible $\ell=5$ single particle component, and so will overwhelmingly $\gamma$ decay. For the state at 5442 keV, tentatively assigned $(5/2^+)$ with an $\ell=2$ centrifugal barrier, $\gamma$-ray and neutron emission will compete in the depopulation of the level as will be discussed in Sect. \ref{sec:level6}.

The population fraction $b_f$ of each level in the reaction, also indicated in Fig. \ref{LS}, is obtained from integrating the $\gamma$-ray peak intensities, corrected for efficiency and feeding of the state by other transitions. In this scheme of $\gamma$-ray tagging, the population of the ground state is obtained by subtraction of the total excited-state cross section from the inclusive cross section. The inclusive cross section was determined from the total number of $^{33}$Si knockout residues divided by the product of  the numbers of incoming projectiles and  target atoms and was measured to be $\sigma^{inc}_{exp}$ = 116(6)~mb. The present value is in agreement with the value of 123(14) mb obtained by Enders {\it et al.} \cite{Ende12} using the same target material and a slightly lower beam energy. The contribution of the higher-lying states (2.8\%, with corresponding transitions shown in red in Fig.~\ref{LS}), was subtracted to obtain a direct knockout cross section of 113(6) mb.

It is important to note the following points. First, the $b_f$ value of the $7/2^-$ isomeric state could not be measured with GRETINA as its lifetime is too long to be detected in-flight. Therefore, it was deduced indirectly, as discussed at the end of Sect.~\ref{sec:level4A} and shown in the inset of Fig. \ref{momenta} g. Secondly, very weakly populated states and transitions could be missing from the level scheme presented in Fig. \ref{LS}. As the sensitivity limit of the present experiment is about $b_f \leqslant1\%$, one
cannot exclude that a few unobserved transitions below this intensity limit exist and, therefore, the ground-state population is slightly overestimated. With a higher sensitivity limit, Enders {\it et~al.} \cite{Ende12} did not observe several of the transitions reported here. The reported ground-state population of 54~(13)\% is consistent with the value, 50.8~(55)\%, reported here. Third, as discussed above, the deduced unbound 5442~keV state population represents a lower limit, as it is deduced only from its $\gamma$-decay component. The relative intensities were computed normalized to the 1010 keV transition. The obtained relative intensities for the 971 keV, 1010 keV, 1724 keV, 1773 keV, 4268 keV, 4347 keV and 5442 keV transitions are 12.6(41)\%, 100(8)\%, 10.6(39)\%, 7.54(60)\%, 2.67(24)\%, 22.6(45)\% and  1.16(98)\% respectively.

\section{\label{sec:level3} Knockout reaction model calculation}

One-nucleon removal reactions from intermediate-energy beams have been used extensively as tools to deduce unique spectroscopic information on the dominant proton and neutron single-particle structures of rare isotopes~\cite{hansen03}. In such one-nucleon removal reactions, a single nucleon, here a neutron, is removed in the grazing collision of a fast-moving projectile, here $^{34}$Si, with a light target nucleus, here $^9$Be. The number and momenta of the fast, forward-traveling $^{33}$Si residual nuclei encode the nuclear structure information of interest. The cross sections to
individual final states scale with the number of neutrons available to be removed from the relevant orbital and thus offer the opportunity to extract spectroscopic factors by comparisons to reaction theory. The shape of the parallel momentum distribution of the knockout residues is sensitive to the $\ell$-value of the orbital in the ground state of the $^{34}$Si projectile from which the neutron is removed. Since this ground state has spin-parity $0^+$, the final state in $^{33}$Si characterizes the neutron orbital involved.

As is usual, the spectroscopic factor for a $^{33}$Si final state, $f$, resulting from the removal of a neutron with quantum numbers $n,\ell,j$, are deduced from the measured, $\sigma_{exp}^f$ (the fourth column of Table~\ref{tab:tabexp}), and theoretical, ${\sigma_{sp}}$, partial cross sections, using
\begin{equation}
C^2 S_{exp} = {{b_f} \sigma^{inc}_{exp}}/{\sigma_{sp}},
\end{equation}
where $\sigma_{exp}^{inc}$ is the measured inclusive knockout cross section and $b_f$ denotes the population fraction for the final state $f$ as determined from $\gamma$-ray spectroscopy. The single-particle cross sections $\sigma_{sp}(n,\ell,j, S_n+E_f)$ are calculated assuming eikonal reaction dynamics \cite{Toste14}. The summed spectroscopic factors $C^2 S_{exp}$, that in a sum-rule limit relate to the occupancy of the orbit, have maximum value $(2j+1)$ in the case of a fully occupied orbital with angular momentum $j$.

The $\sigma_{sp}$ (see e.g. Ref.~\cite{Gade08} for details), the sum of cross sections from the stripping and diffraction mechanisms, are computed from the residue- and neutron-target elastic eikonal $S$ matrices using the double- and single-folding optical limit of Glauber's multiple-scattering theory, respectively. The residue-target interaction uses the proton and neutron densities of $^{33}$Si from SkX Skyrme Hartree-Fock (HF) calculations~\cite{Skyrme}. The $^9$Be target density is assumed of Gaussian shape with a $rms$ radius of 2.36~fm. The bound neutron-$^{33}$Si orbital wave functions and their single-particle $rms$ radii being constrained, consistently, by the corresponding HF calculations \cite{Gade08}.

\section{\label{sec:level4}RESULTS AND DISCUSSION}
\subsection{\label{sec:level4A}Momentum distributions}

The shapes of the longitudinal momentum distributions, $p_\|$, of the one-nucleon knockout residues depend on the $\ell$ value of the removed neutron~\cite{hansen03,bertulani04}. The theoretical momentum distributions were calculated following Ref. \cite{MOMDIS} using the same elastic $S$ matrices, etc. as were used for the computation of the $\sigma_{sp}$. Comparisons of these calculated and the measured momentum distributions are used to characterize the observed $^{33}$Si final states.

Some words of caution are in order in making such comparisons.
First, the often-observed low-momentum tail of the intermediate-energy momentum distributions, attributed to energy transfers and dissipation in the collisions with the target nuclei that are not explicitly included in the reaction kinematics of the eikonal model calculation (see e.g. \cite{Stro14}). Hence, only the high-momentum part of these distributions will be used for the quantitative comparison.

\begin{figure}
\includegraphics[width=\columnwidth]{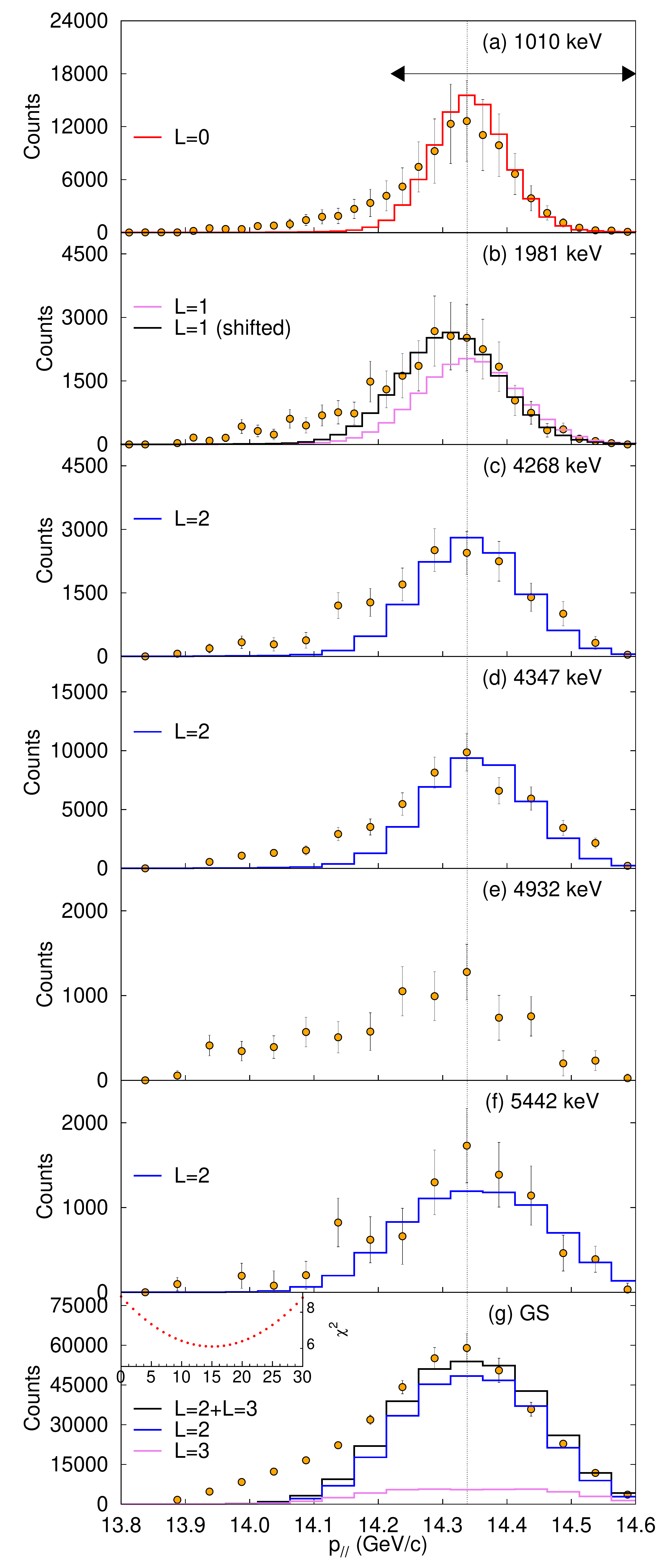}
\caption{(Color online) panels (a) to (f) : Experimental momentum distributions and calculated one-neutron    knockout momentum distributions for individual excited states  states of $^{33}$Si. The distributions are fitted between 14.22  to 14.6 MeV/c (delimited by arrows on the top panel) so as to exclude the low-momentum tail whose shape cannot be described by the eikonal model. (g) Experimental momentum distributions for the ground state of $^{33}$Si along with best fit of the calculation of one neutron knockout with L=2 and L=3 components. The inset shows the $\chi^2$ minimization as function of the fraction of the L=3 component.}
\label{momenta}
\end{figure}

To account for experimental conditions, the theoretical shapes were folded with the momentum profile of the unreacted projectile beam passing through the target to account for the range of momenta originating from the spread of the incoming beam and the momentum straggling in the target. Second, as the higher-spin states, $(9/2^-)$ and $(11/2^-)$, cannot be populated directly from the $^{34}$Si ground state in a
one-step, neutron knockout, such a comparison is not possible. The distribution for the highest populated $11/2^-$ state at 4932 keV is shown for information in Fig. \ref{momenta}(e). Third, for states which are also populated from higher-lying discrete levels, the feeding contributions from the higher-lying states were removed by subtracting its $p_\|$ distribution, scaled by the relative efficiency for observation of the two $\gamma$ rays of the corresponding cascade. Finally, the ground-state distribution was obtained by subtracting the distributions of all states observed feeding it. Since decay of the $7/2^-$ isomer could not be directly measured and subtracted, the experimental ground-state momentum distribution includes the contribution from the $7/2^-$ isomer, if populated (see Fig. \ref{momenta}).

The state at 1010 keV is fed from above by the 971 keV $\gamma$ ray, whose contribution was subtracted. The  $p_\|$ distribution corresponding to the 1010~keV state, shown in Fig.~\ref{momenta}(a), is consistent with an $\ell$=0 neutron removal from the $1s_{1/2}$ orbital, leading to a $1/2^+$ assignment, in agreement with Ref.~\cite{Ende12}. The momentum distribution corresponding to the 1981 keV state, which decays by the 971 keV $\gamma$-ray, seems consistent with an $\ell$=1 neutron removal from the  $1p_{3/2}$ valence orbital. However, the centroid of the distribution is shifted by 0.1 GeV/c to lower momenta, suggesting that the $3/2^-$ state may not be exclusively populated by direct neutron knockout (see later discussion for details). The $\gamma$-gated $p_\|$ distributions of the 4268, 4347 and 5442 keV states agree with both $\ell$=1 and $\ell$=2 values, with a slight preference of the $\chi^2$ fit for the $\ell$=2 assignment, as shown in Fig.~\ref{momenta} (c), (d) and (f), respectively. As no strong $\ell$=1 contribution is expected in this energy range (which would correspond to significant additional $p$-wave strength), we propose that these states arise from the removal of an $\ell$=2 neutron from the $0d_{5/2}$ orbital, leading to the corresponding $5/2^+$ hole states in $^{33}$Si.

A two-component fit of the experimental $p_\|$ distribution of the ground state, including $\ell$=2 and $\ell$=3 contributions, is shown in Fig. \ref{momenta}(g). The best fit was obtained for a $\ell$=3 contribution of 15(9)\%. The $\ell$=2 assignment for the ground state is consistent with its $3/2^+$ configuration, corresponding to a hole in the neutron $0d_{3/2}$ orbital. The additional (weak) $\ell$=3 fraction likely corresponds to a contribution from the $7/2^-$ isomeric state at 1435 keV whose $\gamma$-ray decay is delayed and, therefore, could not be observed with prompt $\gamma$-ray spectroscopy at the target location.

\subsection{\label{sec:level5} Spectroscopic Factors}

The occupancy of an orbital with quantum numbers $n,\ell,j$ from nucleon removal reactions is derived from the sum of the $C^{2}S_{exp}$ values corresponding to the $j^\pi$ final states. The challenge of determining absolute spectroscopic factors/occupancies is long standing. From a measurement point of view, it is experimentally difficult to collect all of the $C^{2}S_{exp}$ strength, especially in removals of deeply-bound orbitals that in general lead to both particle-bound and particle-unbound final states, so the occupancy values derived rarely reach the maximum for a given orbital. Related to this, there is evidence of the need for a quenching of deduced $C^{2}S_{exp}$, by of order 0.6-0.7, relative to independent-particle model expectations from $(e,e'p)$ reaction measurements on stable nuclei (see e.g. Refs. \cite{Dick04,Dugu15}). This effect is generally attributed to short- and longer-range correlations that fragment the spectroscopic strengths to higher energy/momenta and orbitals, with the result that a fraction of strength lies: (a) beyond experimental detection, and (b) typically outside the truncated model spaces of shell-model calculations. For rare isotopes studied using $^9$Be-target-induced knockout reactions, as used here, it is observed \cite{Toste14} that a reduction factor $R_s=\sigma_{exp}^{inc}/ \sigma_{th }^{inc}$, derived as the ratio of the measured inclusive
cross-section to the calculated eikonal model cross-section to all bound shell-model final states, is dependent on the asymmetry of the projectile's proton and neutron separation energies. That is, that the reaction model used here, combined with shell model spectroscopic factors, predicts too much cross section to bound final states. From the systematics of \cite{Toste14}, one would expect $R_s=0.8(2)$ when using an appropriate shell-model in the present case. This will be discussed in the context of the shell-model calculations presented in Section \ref{sec:level7}. In the $N=Z$, $^{40}$Ca$(p,d)$$^{39}$Ca transfer reaction case of Ref. \cite{Mato93}, involving the isotone of $^{33}$Si, the normalization was achieved so that the summed number of the neutrons in the $1s_{1/2}$, $0d_{3/2}$ and $0f_{7/2}$ orbits was equal to 6. Thus, despite the asymmetry and reaction mechanism differences in the two experimental approaches, the normalizations appear approximately equivalent. We will discuss and compare occupancy values for the two works.

As shown in the second to last column of Table~\ref{tab:tabexp}, the spectroscopic factor deduced for the ground state, $C^{2}S_{exp} $=3.90(47), corresponds to a fully occupied $0d_{3/2}$ orbital, which can hold up to 4 neutrons, with an expectation that no other $3/2^+$ state with significant strength is present. A similar value of 3.74(20) was obtained for neutron removal to the ground state of $^{39}$Ca \cite{Mato93} using the $(p,d)$ neutron pickup reaction. Taken at face value, this suggests that the $^{40}$Ca and $^{34}$Si closed-shell neutron cores may be similar.

The value obtained for the $1/2^+$ state, 1.34(8), is lower than the full occupancy of the $1s_{1/2}$ orbital, and smaller than the value of 1.64(15) for $^{39}$Ca. This smaller $C^{2}S_{exp}$ value suggests either the presence of another $1/2^+$ state at higher excitation energy and/or the depletion of the $1s_{1/2}$ orbital through population of the valence $fp$ shell. We found no evidence of a second, bound $1/2^+_2$ state, which may be neutron unbound. As to the second hypothesis, a possible migration of neutrons to the upper $fp$ orbits, one has to ask why this would occur more significantly from $1s_{1/2}$ rather than from the less deeply-bound $0d_{3/2}$ orbit, making this explanation less likely.

Being located well below the Fermi surface, the  $0d_{5/2}$ strength is expected to be fragmented and hence not observed in a single state. Three $(5/2^+)$ states are proposed, corresponding to a summed  $C^{2}S_{exp}$ value of 0.91(5). As the 5442 keV state is unbound, its deduced value of 0.10(3), from its $\gamma$ decay branch, is a lower limit. As compared to the case of $^{39}$Ca \cite{Mato93}, the summed $5/2^+$ strength is of similar magnitude, at about 5.1 MeV.

We now focus the discussion on the valence states. The $C^{2}S_{exp}$-values for the $7/2^-$ (0.72(43)) and $3/2^-$ (0.35(5)) states indicate a rather large neutron occupation of the $0f_{7/2}$ and $1p_{3/2}$ valence orbits in $^{34}$Si. As such, this finding is hardly compatible with the almost full occupancy of the $0d_{3/2}$ and the non negligible occupancy of the $1s_{1/2}$ orbital. For comparison, smaller $C^2S$ values have been found for the valence states in $^{39}$Ca : 0.14 for the $7/2^-$ and 0.01 for the $3/2^-$ states. As will be shown in Sect.~\ref{sec:level7}, shell-model calculations agree very well with the $C^{2}S_{exp}$ of the $3/2^+$ and $1/2^+$ states, but predict significantly smaller occupation numbers for the $7/2^-$ and $3/2^-$ states, in agreement with what was reported for the case of $^{39}$Ca.

\subsection{\label{sec:level6} Discussion on the production of negative-parity states}

Direct, one-step neutron knockout from the occupied $sd$-shell orbits in $^{34}$Si will
populate positive-parity final states. Depending on the fraction of neutron pairs scattered
from the $sd$-shell to the valence $1p_{3/2}$ and $0f_{7/2}$ orbits, $3/2^-$ and $7/2^-$ 
final states may also be directly populated. However, the population of negative-parity
states with spin and parity $(11/2^-)$, spins supported by the observation of their decay
branches, is also observed in $^{33}$Si with a branch, $b_f$, of up to a few percent. This is 
indicative of a higher-order reaction mechanism contribution. 

Such a contribution was also observed and discussed in the $^{36}$S(-1p) and $^{34}$Si(-1p)
proton knockout experimental studies of Mutschler et al. \cite{Muts16b,Muts16}. Similarly,
positive-parity states populated from $sd$-shell knockouts were expected, exclusively. There,
negative-parity states were identified in $^{35}$P comprising 4.7(9) \% of the inclusive
cross section. The centroids of the residue momentum distributions for these states were
also shifted to lower values, pointing to production by another mechanism. In the 
$^{34}$Si(-1p) experiment \cite{Muts16}, the deduced feeding of negative-parity states was
7.7(11) \% \cite{MutsPHD}.

In the current work, Fig. \ref{LS}, the measured quantitative indication of a higher-order 
contribution is the approximately 2.8 \% of the intensity attributed to the 11/2$^-$ negative-parity states. Based on the other available experiments, discussed above, one might speculate 
that an additional fraction of $\simeq$ 4-5 \% of the cross section could involve the population 
of other states by indirect mechanisms, including 7/2$^-$ and/or 3/2$^-$ states. The deduced 
$b_f$ value for the 7/2$^-$ isomeric state, 9.0(54) \%, is very uncertain and compatible with 
zero at less than 2$\sigma$. Further, as the momentum distribution of the $7/2^-$ isomer could 
not be disentangled from that of the ground state, we cannot determine whether the $7/2^-$ 
isomer population is in part due to higher-order mechanisms. The percentage population for 
the 3/2$^-$ state, of 4.5(6) \%, is clearly of the same order as observed higher-order 
contributions. Such a component would be consistent with the observed shift in the $3/2^-$ 
state momentum distribution centroid to lower momentum relative to the positive-parity states, 
as is shown in Fig. \ref{momenta}(b) and discussed in Sect.~\ref{sec:level4A}.

So, one may speculate that a fraction, perhaps as much as half, of the small occupancies 
of the $fp$-shell orbits deduced assuming only a direct knockout mechanism, are actually the 
result of undetermined, higher-order pathways. We comment that the experiments cited above 
\cite{Muts16, Muts16b} were proton knockout reactions, while the current work is a neutron 
knockout reaction. Thus, depending on the higher-order mechanism(s) involved, estimated 
contributions could differ.

In concluding this discussion, we note also that in recent one-proton and one-neutron removal
studies on $^{56}${Ni}, core-coupled final states in $^{55}$Ni and $^{55}$Co with complex 
structures, of $2^+_1$~($^{56}$Ni) ~$\otimes f_{7/2}^{-1}$ character, were observed to be 
weakly populated, having shifted momentum distributions~\cite{spieker19}. This suggests 
mechanisms involved in the high-spin states population may include nucleon knockout from 
an excited state of the projectile \cite{spieker19,Stro14,Muts16b}. Continued experimental 
advances from earlier, more inclusive data limitations to higher-statistics, final-states-exclusive 
cross sections and, importantly, their momentum distributions thus begin to allow such 
contributions to be identified and quantified. This is important to better determine
the (dominant) direct knockout contributions and the associated deduced single-particle
spectroscopy.

\subsection{\label{sec:level7} Comparison with Shell Model Calculations}

Shell model calculations have been performed using the $sdpf$ valence space in order to predict the energy, spin and $C^2S$ values of the levels populated in $^{33}$Si from the neutron knockout reaction. The effective interaction used is a refinement of the {\sc sdpf-u-mix} interaction \cite{SDPF-UMIX}. In particular, the proton-neutron monopole interaction incorporates more constraints on details of the underlying single particle behavior in order to reproduce the observed $3/2^-$  and  $7/2^-$  crossing and location of the $pf$ intruder orbitals in $^{27}$Ne, $^{29}$Mg and $^{31}$Si. In the meantime, the neutron-neutron monopole interaction has been adjusted to leave the physics of the N=20 island of inversion unchanged.

As shown in Table~\ref{tab:tabSM}, most of the predicted states (except the $1/2_2^+$ and $3/2_2^+$) are observed and their ordering is correct. The energies of the first two excited states are predicted to be slightly too low: the $1/2^+$ is calculated at 750 keV, compared to 1010 keV, and the $7/2^-$ at 1.35 MeV, compared to 1.435 MeV. The calculated and experimental energies of the $3/2^-$ agree within 60 keV, while the first calculated $5/2^+$ state, with a significant $C^2S$ value, appears at 4.29 MeV, comparable to the value of 4.347 MeV reported in the experiment.

The calculated ground state $C^2S$=3.33 value is smaller than that deduced experimentally, 3.90(47). However, adding the $3/2^+_2$ state strength, not observed experimentally, the summed spectroscopic strength of the $0d_{3/2}$ orbit is 3.5, close to the experimental value. This $3/2^+_2$ has not been observed experimentally. Either it has a much weaker $C^2S$ value, which would indicate that the $3/2^+$ strength is less fragmented than is calculated, or/and the $3/2^+_2$ state decays through various transitions, reducing its chance to be observed.

The calculated $C^2S$ of the $1/2^+$ state, 1.46, is in relatively good agreement with the experimental value of 1.34(8). Adding the $C^2S$ value of 0.47 of the $1/2^+_2$ at 4.4 MeV leads to a summed spectroscopic strength of 1.93, close to the sum rule value of 2. The  $1/2^+_2$ state is predicted to lie just 100 keV below the neutron emission threshold. It has not been observed experimentally, which may mean this $\ell$=0 state is located above and not below the neutron threshold, decaying by neutron emission.

\begin{table}
\caption{\label{tab:tabSM} Comparison of experimental and shell-model calculations of energies and
$C^2S$ values. The theoretical partial cross sections to the bound shell model final states are
also shown.}
\begin{ruledtabular}
\begin{tabular}{cccccc}
$J^{\pi}$  & $E_{exp}$  & $E_{th}$ &$C{^2}S_{exp} $ &  $C{^2}S_{th}$&$\sigma_{th}$\\
&$[MeV]$& $[MeV]$ &  &&[mb]\\
\hline
$3/2^+$ & 0 & 0 &   3.90(47)&   3.33                        &53.50  \\
$1/2^+$ & 1.010 & 0.75 &  1.34(8) &   1.46                  &29.49  \\
$7/2^-$ & 1.435 & 1.35 &  0.72(43) \footnotemark[1] & 0.13  &1.88  \\
$3/2^-$ & 1.981 & 1.92 &  0.35(5) \footnotemark[1]&   0.015 &0.23 \\
$3/2^+_2$ &  & 3.27 &   &   0.17                            &2.12  \\
$5/2^+$ & 4.268 & 3.90 &  0.15(3) &   0.067                 &0.91  \\
$5/2^+$ & 4.347 & 4.29 &  0.66(6) &   2.24                  &29.73  \\
$1/2^+_2$ &  & 4.41 &   &   0.47                            &7.18  \\
$5/2^+$ & 5.442 &  5.09 &  0.10(3) &   0.24& \\
$5/2^+$ & -        &  5.81 &   &   0.40& \\
\end{tabular}
\end{ruledtabular}
\footnotetext[1]{Values may be overestimated due to a contributions from higher-order
processes (see discussion in Sect.\ref{sec:level6}).}
\end{table}

The theoretical $C^2S$ values of the $7/2^-$ and $3/2^-$ states, 0.13 and 0.015, respectively, are significantly smaller than the experimental ones. As discussed in the previous section, this  supports the suggestion that part of the population of these weaker states is by a higher-order reaction mechanism.

The three $5/2^+$ states reported here, one of which is about 940 keV above $S_n$, all decay directly to the $3/2^+$ ground state. This decay pattern suggests that these states decay by $M1$ transitions, as is expected between spin-orbit partner states. Three $5/2^+$ states are also calculated up to 5 MeV, among which one state carries a large $C^2S$ value of 2.24. Their summed value is 2.55 up to 5.1 MeV, almost half of the expected occupancy of the $0d_{5/2}$ orbital. Experimentally, the summed value reaches only 0.91(5). Unfortunately, as only a lower limit can be obtained for the observed unbound state at 5.442 MeV, a more quantitative comparison with theory is not possible. Such a comparison would have provided useful information on the amplitude of the neutron $0d_{3/2} - 0d_{5/2}$ spin-orbit splitting and its comparison to theory. To assess the possible reduction of the neutron $0d_{3/2}-0d_{5/2}$ splitting between $^{39}$Ca and $^{33}$Si, another experimental technique, such as $^{34}$Si$(p,d)$$^{33}$Si, must be used to identify the unbound $5/2^+$ states in $^{33}$Si.

One can, however, estimate the fraction of the 5.442 MeV decay that proceeds by neutron decay, rather than $\gamma$-ray emission, to constrain the fraction of $C^2S$ value missing for the 5.442 MeV state. A partial width of 180 keV, or 2.7$\cdot$10$^{20}$ s$^{-1}$, is calculated using a Woods Saxon potential and an $\ell$=2 neutron tunneling through the centrifugal barrier.  This value should be multiplied by the $C^{2}S$ value for the decay of this state to the $^{32}$Si ground state, which is of the order of 10$^{-3}$ (with a rather large uncertainty). Taking this value, one obtains a partial neutron decay lifetime of 2.7$\cdot$10$^{17}$ s$^{-1}$. The $B(M1)$ values corresponding to the decays of the three $5/2^+$ states are 0.176, 0.225 and 0.063 $\mu_N^2$ at 3.4, 3.69 and 5.0 MeV, respectively. Taking the fastest calculated $B(M1)$ value of 0.22 $\mu_N^2$, it corresponds to 2$\cdot$10$^{14} $ s$^{-1}$. Thus, the neutron emission is expected to be much faster than the $\gamma$-decay, by about 3 orders of magnitude. Given this dominance of the neutron decay branch, it follows that $C^2S$ to the 5.442 MeV level must be large in order to allow observation of its $\gamma$-ray partial decay branch.

Finally, given the shell model calculations of Table \ref{tab:tabSM}, we can also now compute the quenching factor $R_s$ for this case. That is, the ratio of the measured and theoretical inclusive knockout cross sections to states up to the $S_n$ value of 4.5 MeV.  The experimental value is derived from the observed cross section $\sigma^{inc}_{exp}$=116(6) mb to all observed states, from which we subtract: (i) 3.2(4) mb from the feeding of the two 11/2$^-_{1,2}$ states, due to a higher-order mechanism, and (ii) 1.25(35) mb from the feeding of the unbound $5/2^+$ state at 5.44 MeV. In the absence of further diagnostics, we neglect possible contributions from higher-order processes to the $7/2^-$ and  $3/2^-$ states. The cross section to these two states (Table~\ref{tab:tabexp}) is 15.6(6) mb. The resulting experimental cross section is $111\pm 7 $ mb.  With the calculated inclusive knockout cross section, from Table~\ref{tab:tabSM}, of $\sigma^{inc}_{th}$=125 mb, we obtain $R_s=0.89(6)$, which is consistent with the systematics for other systems \cite{Toste14}. The overall effective neutron separation energy, derived by weighting the separation energy to each bound shell-model final state by its partial cross section, is 9.07 MeV, and hence the neutron-proton separation energy asymmetry \cite{Gade08} is $\Delta S=-9.71$ MeV. It is of considerable interest to understand the origins of this $R_s$ suppression, the detail of which may differ for each projectile and reaction. In this case, and with the presented shell-model calculation, the measured and calculated $3/2^+$ and $1/2^+$ cross sections are in broad agreement and, primarily, the $R_s$ suppression is the result of significant additional shell-model spectroscopic strength generating large cross sections to bound $5/2^+$ final states, 30.6 mb theoretically, compared to the measured value of 10.8 mb. In the absence of detailed spectra, as available here, $R_s$ has sometimes been applied as an overall renormalization of the calculated partial cross sections prior to the extraction of experimental spectroscopic factors. Our analysis gives no justification for using $R_s$ in this way for the $^{33}$Si case and this was therefore not done in Section \ref{sec:level3}.

\section{Conclusions}
The one-neutron knockout reaction $^9$Be($^{34}$Si, $^{33}$Si $+ \gamma$)X  has been used to populate excited states in $^{33}$Si, with the goal of probing the Fermi surface of $^{34}$Si. Spectroscopic factors of orbits bounding the $N=20$ gap were derived from the partial cross sections to individual states, mostly obtained from the detection of their $\gamma$-ray transitions with the GRETINA array. Spin assignments were proposed by comparing the measured, exclusive parallel momentum distribution to calculated ones using eikonal reaction dynamics. As for the normally occupied states below the Fermi surface, the $C^2S$ values of the $3/2^+$ ground state and $1/2^+$ first-excited state agree reasonably well with shell-model calculations, pointing to a strong $N=20$ shell closure. Two new states with $(5/2^+)$ assignment, likely corresponding to a neutron hole in the more deeply-bound $0d_{5/2}$ orbital, are identified, one of which lies about 1 MeV above the neutron emission threshold. Owing to its likely high partial neutron decay probability, a direct comparison  between experimental and calculated $(5/2^+)$ strength is not straightforward.

Thanks to the sensitivity of GRETINA, this experiment has also reconfirmed the weak population of states (with up to a few percent of the inclusive cross section) which cannot originate from a single-step one-neutron removal from the projectile ground state. In particular, the population of two high-$j$ negative-parity states with $11/2^-$ assignment is clearly evidenced through their decays to the lower-lying $9/2^-$ and $7/2^-$ states. This observation, together with a downshift observed for the $3/2^-$ parallel momentum distribution and the lack of the $7/2^-$ momentum distribution for similar diagnostics, casts uncertainty on the population of the other negative-parity states, $7/2^-$  and $3/2^-$, as resulting from a pure knockout reaction from partially filled  $0f_{7/2}$ and $1p_{3/2}$ orbits, respectively. We, therefore, caution that, when aiming at a percentage level of precision in the determination of weaker spectroscopic factors from knockout reactions, sufficient statistics and momentum resolution are required for the extraction of accurate exclusive momentum distributions that allow the tell-tale identification of weak, presumably higher-order processes from a downshift in these distributions.

\section*{ACKNOWLEDGMENTS}
We thank F.~De~Oliveira for providing the calculation of the neutron decay width. This work is supported by the National Science Foundation (NSF) under Grant Nos. PHY-1102511 and PHY-1306297, the Institut Universitaire de France, and by the National Research Foundation of South Africa under Grant No. 118846. GRETINA was funded by the US DOE - Office of Science. Operation of the array at NSCL is supported by NSF under  Cooperative Agreement PHY-1102511 (NSCL) and DOE under grant DE-AC02-05CH11231  (LBNL). J.A.T acknowledges support of the Science and Technology Facility Council (UK) grant ST/L005743/1.

\bibliographystyle{apsrev4-1}
\bibliography{34KOV8.bib}

\begin{thebibliography}{42}%
\makeatletter
\providecommand \@ifxundefined [1]{%
 \@ifx{#1\undefined}
}%
\providecommand \@ifnum [1]{%
 \ifnum #1\expandafter \@firstoftwo
 \else \expandafter \@secondoftwo
 \fi
}%
\providecommand \@ifx [1]{%
 \ifx #1\expandafter \@firstoftwo
 \else \expandafter \@secondoftwo
 \fi
}%
\providecommand \natexlab [1]{#1}%
\providecommand \enquote  [1]{``#1''}%
\providecommand \bibnamefont  [1]{#1}%
\providecommand \bibfnamefont [1]{#1}%
\providecommand \citenamefont [1]{#1}%
\providecommand \href@noop [0]{\@secondoftwo}%
\providecommand \href [0]{\begingroup \@sanitize@url \@href}%
\providecommand \@href[1]{\@@startlink{#1}\@@href}%
\providecommand \@@href[1]{\endgroup#1\@@endlink}%
\providecommand \@sanitize@url [0]{\catcode `\\12\catcode `\$12\catcode
  `\&12\catcode `\#12\catcode `\^12\catcode `\_12\catcode `\%12\relax}%
\providecommand \@@startlink[1]{}%
\providecommand \@@endlink[0]{}%
\providecommand \url  [0]{\begingroup\@sanitize@url \@url }%
\providecommand \@url [1]{\endgroup\@href {#1}{\urlprefix }}%
\providecommand \urlprefix  [0]{URL }%
\providecommand \Eprint [0]{\href }%
\providecommand \doibase [0]{http://dx.doi.org/}%
\providecommand \selectlanguage [0]{\@gobble}%
\providecommand \bibinfo  [0]{\@secondoftwo}%
\providecommand \bibfield  [0]{\@secondoftwo}%
\providecommand \translation [1]{[#1]}%
\providecommand \BibitemOpen [0]{}%
\providecommand \bibitemStop [0]{}%
\providecommand \bibitemNoStop [0]{.\EOS\space}%
\providecommand \EOS [0]{\spacefactor3000\relax}%
\providecommand \BibitemShut  [1]{\csname bibitem#1\endcsname}%
\let\auto@bib@innerbib\@empty
%</preamble>
\bibitem [{\citenamefont {Grasso}\ \emph {et~al.}(2009)\citenamefont {Grasso},
  \citenamefont {Gaudefroy}, \citenamefont {Khan}, \citenamefont {Nik\ifmmode
  \check{s}\else \v{s}\fi{}i\ifmmode~\acute{c}\else \'{c}\fi{}}, \citenamefont
  {Piekarewicz}, \citenamefont {Sorlin}, \citenamefont {Giai},\ and\
  \citenamefont {Vretenar}}]{Grass09}%
  \BibitemOpen
  \bibfield  {author} {\bibinfo {author} {\bibfnamefont {M.}~\bibnamefont
  {Grasso}}, \bibinfo {author} {\bibfnamefont {L.}~\bibnamefont {Gaudefroy}},
  \bibinfo {author} {\bibfnamefont {E.}~\bibnamefont {Khan}}, \bibinfo {author}
  {\bibfnamefont {T.}~\bibnamefont {Nik\ifmmode \check{s}\else
  \v{s}\fi{}i\ifmmode~\acute{c}\else \'{c}\fi{}}}, \bibinfo {author}
  {\bibfnamefont {J.}~\bibnamefont {Piekarewicz}}, \bibinfo {author}
  {\bibfnamefont {O.}~\bibnamefont {Sorlin}}, \bibinfo {author} {\bibfnamefont
  {N.~V.}\ \bibnamefont {Giai}}, \ and\ \bibinfo {author} {\bibfnamefont
  {D.}~\bibnamefont {Vretenar}},\ }\href {\doibase 10.1103/PhysRevC.79.034318}
  {\bibfield  {journal} {\bibinfo  {journal} {Phys. Rev. C}\ }\textbf {\bibinfo
  {volume} {79}},\ \bibinfo {pages} {034318} (\bibinfo {year}
  {2009})}\BibitemShut {NoStop}%
\bibitem [{\citenamefont {Mutschler}\ \emph {et~al.}(2017)\citenamefont
  {Mutschler}, \citenamefont {Lemasson}, \citenamefont {Sorlin}, \citenamefont
  {Bazin}, \citenamefont {Borcea}, \citenamefont {Borcea}, \citenamefont
  {Dombr\'adi}, \citenamefont {Ebran}, \citenamefont {Gade}, \citenamefont
  {Iwasaki}, \citenamefont {Khan}, \citenamefont {Lepailleur}, \citenamefont
  {Recchia}, \citenamefont {Roger}, \citenamefont {Rotaru}, \citenamefont
  {Sohler}, \citenamefont {Stanoiu}, \citenamefont {Stroberg}, \citenamefont
  {Tostevin}, \citenamefont {Vandebrouck}, \citenamefont {Weisshaar},\ and\
  \citenamefont {Wimmer}}]{Muts16}%
  \BibitemOpen
  \bibfield  {author} {\bibinfo {author} {\bibfnamefont {A.}~\bibnamefont
  {Mutschler}}, \bibinfo {author} {\bibfnamefont {A.}~\bibnamefont {Lemasson}},
  \bibinfo {author} {\bibfnamefont {O.}~\bibnamefont {Sorlin}}, \bibinfo
  {author} {\bibfnamefont {D.}~\bibnamefont {Bazin}}, \bibinfo {author}
  {\bibfnamefont {C.}~\bibnamefont {Borcea}}, \bibinfo {author} {\bibfnamefont
  {R.}~\bibnamefont {Borcea}}, \bibinfo {author} {\bibfnamefont
  {Z.}~\bibnamefont {Dombr\'adi}}, \bibinfo {author} {\bibfnamefont {J.~P.}\
  \bibnamefont {Ebran}}, \bibinfo {author} {\bibfnamefont {A.}~\bibnamefont
  {Gade}}, \bibinfo {author} {\bibfnamefont {H.}~\bibnamefont {Iwasaki}},
  \bibinfo {author} {\bibfnamefont {E.}~\bibnamefont {Khan}}, \bibinfo {author}
  {\bibfnamefont {A.}~\bibnamefont {Lepailleur}}, \bibinfo {author}
  {\bibfnamefont {F.}~\bibnamefont {Recchia}}, \bibinfo {author} {\bibfnamefont
  {T.}~\bibnamefont {Roger}}, \bibinfo {author} {\bibfnamefont
  {F.}~\bibnamefont {Rotaru}}, \bibinfo {author} {\bibfnamefont
  {D.}~\bibnamefont {Sohler}}, \bibinfo {author} {\bibfnamefont
  {M.}~\bibnamefont {Stanoiu}}, \bibinfo {author} {\bibfnamefont {S.~R.}\
  \bibnamefont {Stroberg}}, \bibinfo {author} {\bibfnamefont {J.~A.}\
  \bibnamefont {Tostevin}}, \bibinfo {author} {\bibfnamefont {M.}~\bibnamefont
  {Vandebrouck}}, \bibinfo {author} {\bibfnamefont {D.}~\bibnamefont
  {Weisshaar}}, \ and\ \bibinfo {author} {\bibfnamefont {K.}~\bibnamefont
  {Wimmer}},\ }\href {\doibase 10.1038/nphys3916} {\bibfield  {journal}
  {\bibinfo  {journal} {Nature Physics}\ }\textbf {\bibinfo {volume} {13}},\
  \bibinfo {pages} {152} (\bibinfo {year} {2017})}\BibitemShut {NoStop}%
\bibitem [{\citenamefont {Duguet}\ \emph {et~al.}(2017)\citenamefont {Duguet},
  \citenamefont {Som\`a}, \citenamefont {Lecluse}, \citenamefont {Barbieri},\
  and\ \citenamefont {Navr\'atil}}]{Dugu17}%
  \BibitemOpen
  \bibfield  {author} {\bibinfo {author} {\bibfnamefont {T.}~\bibnamefont
  {Duguet}}, \bibinfo {author} {\bibfnamefont {V.}~\bibnamefont {Som\`a}},
  \bibinfo {author} {\bibfnamefont {S.}~\bibnamefont {Lecluse}}, \bibinfo
  {author} {\bibfnamefont {C.}~\bibnamefont {Barbieri}}, \ and\ \bibinfo
  {author} {\bibfnamefont {P.}~\bibnamefont {Navr\'atil}},\ }\href {\doibase
  10.1103/PhysRevC.95.034319} {\bibfield  {journal} {\bibinfo  {journal} {Phys.
  Rev. C}\ }\textbf {\bibinfo {volume} {95}},\ \bibinfo {pages} {034319}
  (\bibinfo {year} {2017})}\BibitemShut {NoStop}%
\bibitem [{\citenamefont {Burgunder}\ \emph {et~al.}(2014)\citenamefont
  {Burgunder}, \citenamefont {Sorlin}, \citenamefont {Nowacki}, \citenamefont
  {Giron}, \citenamefont {Hammache}, \citenamefont {Moukaddam}, \citenamefont
  {de~S\'er\'eville}, \citenamefont {Beaumel}, \citenamefont {C\`aceres},
  \citenamefont {Cl\'ement}, \citenamefont {Duch\^ene}, \citenamefont {Ebran},
  \citenamefont {Fernandez-Dominguez}, \citenamefont {Flavigny}, \citenamefont
  {Franchoo}, \citenamefont {Gibelin}, \citenamefont {Gillibert}, \citenamefont
  {Gr\'evy}, \citenamefont {Guillot}, \citenamefont {Lepailleur}, \citenamefont
  {Matea}, \citenamefont {Matta}, \citenamefont {Nalpas}, \citenamefont
  {Obertelli}, \citenamefont {Otsuka}, \citenamefont {Pancin}, \citenamefont
  {Poves}, \citenamefont {Raabe}, \citenamefont {Scarpaci}, \citenamefont
  {Stefan}, \citenamefont {Stodel}, \citenamefont {Suzuki},\ and\ \citenamefont
  {Thomas}}]{Burg14}%
  \BibitemOpen
  \bibfield  {author} {\bibinfo {author} {\bibfnamefont {G.}~\bibnamefont
  {Burgunder}}, \bibinfo {author} {\bibfnamefont {O.}~\bibnamefont {Sorlin}},
  \bibinfo {author} {\bibfnamefont {F.}~\bibnamefont {Nowacki}}, \bibinfo
  {author} {\bibfnamefont {S.}~\bibnamefont {Giron}}, \bibinfo {author}
  {\bibfnamefont {F.}~\bibnamefont {Hammache}}, \bibinfo {author}
  {\bibfnamefont {M.}~\bibnamefont {Moukaddam}}, \bibinfo {author}
  {\bibfnamefont {N.}~\bibnamefont {de~S\'er\'eville}}, \bibinfo {author}
  {\bibfnamefont {D.}~\bibnamefont {Beaumel}}, \bibinfo {author} {\bibfnamefont
  {L.}~\bibnamefont {C\`aceres}}, \bibinfo {author} {\bibfnamefont
  {E.}~\bibnamefont {Cl\'ement}}, \bibinfo {author} {\bibfnamefont
  {G.}~\bibnamefont {Duch\^ene}}, \bibinfo {author} {\bibfnamefont {J.~P.}\
  \bibnamefont {Ebran}}, \bibinfo {author} {\bibfnamefont {B.}~\bibnamefont
  {Fernandez-Dominguez}}, \bibinfo {author} {\bibfnamefont {F.}~\bibnamefont
  {Flavigny}}, \bibinfo {author} {\bibfnamefont {S.}~\bibnamefont {Franchoo}},
  \bibinfo {author} {\bibfnamefont {J.}~\bibnamefont {Gibelin}}, \bibinfo
  {author} {\bibfnamefont {A.}~\bibnamefont {Gillibert}}, \bibinfo {author}
  {\bibfnamefont {S.}~\bibnamefont {Gr\'evy}}, \bibinfo {author} {\bibfnamefont
  {J.}~\bibnamefont {Guillot}}, \bibinfo {author} {\bibfnamefont
  {A.}~\bibnamefont {Lepailleur}}, \bibinfo {author} {\bibfnamefont
  {I.}~\bibnamefont {Matea}}, \bibinfo {author} {\bibfnamefont
  {A.}~\bibnamefont {Matta}}, \bibinfo {author} {\bibfnamefont
  {L.}~\bibnamefont {Nalpas}}, \bibinfo {author} {\bibfnamefont
  {A.}~\bibnamefont {Obertelli}}, \bibinfo {author} {\bibfnamefont
  {T.}~\bibnamefont {Otsuka}}, \bibinfo {author} {\bibfnamefont
  {J.}~\bibnamefont {Pancin}}, \bibinfo {author} {\bibfnamefont
  {A.}~\bibnamefont {Poves}}, \bibinfo {author} {\bibfnamefont
  {R.}~\bibnamefont {Raabe}}, \bibinfo {author} {\bibfnamefont {J.~A.}\
  \bibnamefont {Scarpaci}}, \bibinfo {author} {\bibfnamefont {I.}~\bibnamefont
  {Stefan}}, \bibinfo {author} {\bibfnamefont {C.}~\bibnamefont {Stodel}},
  \bibinfo {author} {\bibfnamefont {T.}~\bibnamefont {Suzuki}}, \ and\ \bibinfo
  {author} {\bibfnamefont {J.~C.}\ \bibnamefont {Thomas}},\ }\href {\doibase
  10.1103/PhysRevLett.112.042502} {\bibfield  {journal} {\bibinfo  {journal}
  {Phys. Rev. Lett.}\ }\textbf {\bibinfo {volume} {112}},\ \bibinfo {pages}
  {042502} (\bibinfo {year} {2014})}\BibitemShut {NoStop}%
\bibitem [{\citenamefont {Zegers}\ \emph {et~al.}(2010)\citenamefont {Zegers},
  \citenamefont {Meharchand}, \citenamefont {Shimbara}, \citenamefont {Austin},
  \citenamefont {Bazin}, \citenamefont {Brown}, \citenamefont {Diget},
  \citenamefont {Gade}, \citenamefont {Guess}, \citenamefont {Hausmann},
  \citenamefont {Hitt}, \citenamefont {Howard}, \citenamefont {King},
  \citenamefont {Miller}, \citenamefont {Noji}, \citenamefont {Signoracci},
  \citenamefont {Starosta}, \citenamefont {Tur}, \citenamefont {Vaman},
  \citenamefont {Voss}, \citenamefont {Weisshaar},\ and\ \citenamefont
  {Yurkon}}]{Zege10}%
  \BibitemOpen
  \bibfield  {author} {\bibinfo {author} {\bibfnamefont {R.~G.~T.}\
  \bibnamefont {Zegers}}, \bibinfo {author} {\bibfnamefont {R.}~\bibnamefont
  {Meharchand}}, \bibinfo {author} {\bibfnamefont {Y.}~\bibnamefont
  {Shimbara}}, \bibinfo {author} {\bibfnamefont {S.~M.}\ \bibnamefont
  {Austin}}, \bibinfo {author} {\bibfnamefont {D.}~\bibnamefont {Bazin}},
  \bibinfo {author} {\bibfnamefont {B.~A.}\ \bibnamefont {Brown}}, \bibinfo
  {author} {\bibfnamefont {C.~A.}\ \bibnamefont {Diget}}, \bibinfo {author}
  {\bibfnamefont {A.}~\bibnamefont {Gade}}, \bibinfo {author} {\bibfnamefont
  {C.~J.}\ \bibnamefont {Guess}}, \bibinfo {author} {\bibfnamefont
  {M.}~\bibnamefont {Hausmann}}, \bibinfo {author} {\bibfnamefont {G.~W.}\
  \bibnamefont {Hitt}}, \bibinfo {author} {\bibfnamefont {M.~E.}\ \bibnamefont
  {Howard}}, \bibinfo {author} {\bibfnamefont {M.}~\bibnamefont {King}},
  \bibinfo {author} {\bibfnamefont {D.}~\bibnamefont {Miller}}, \bibinfo
  {author} {\bibfnamefont {S.}~\bibnamefont {Noji}}, \bibinfo {author}
  {\bibfnamefont {A.}~\bibnamefont {Signoracci}}, \bibinfo {author}
  {\bibfnamefont {K.}~\bibnamefont {Starosta}}, \bibinfo {author}
  {\bibfnamefont {C.}~\bibnamefont {Tur}}, \bibinfo {author} {\bibfnamefont
  {C.}~\bibnamefont {Vaman}}, \bibinfo {author} {\bibfnamefont
  {P.}~\bibnamefont {Voss}}, \bibinfo {author} {\bibfnamefont {D.}~\bibnamefont
  {Weisshaar}}, \ and\ \bibinfo {author} {\bibfnamefont {J.}~\bibnamefont
  {Yurkon}},\ }\href {\doibase 10.1103/PhysRevLett.104.212504} {\bibfield
  {journal} {\bibinfo  {journal} {Phys. Rev. Lett.}\ }\textbf {\bibinfo
  {volume} {104}},\ \bibinfo {pages} {212504} (\bibinfo {year}
  {2010})}\BibitemShut {NoStop}%
\bibitem [{\citenamefont {Lic\ifmmode~\u{a}\else \u{a}\fi{}}\ \emph
  {et~al.}(2019)\citenamefont {Lic\ifmmode~\u{a}\else \u{a}\fi{}},
  \citenamefont {Rotaru}, \citenamefont {Borge}, \citenamefont {Gr\'evy},
  \citenamefont {Negoi\ifmmode \mbox{\c{t}}\else \c{t}\fi{}\ifmmode~\u{a}\else
  \u{a}\fi{}}, \citenamefont {Poves}, \citenamefont {Sorlin}, \citenamefont
  {Andreyev}, \citenamefont {Borcea}, \citenamefont {Costache}, \citenamefont
  {De~Witte}, \citenamefont {Fraile}, \citenamefont {Greenlees}, \citenamefont
  {Huyse}, \citenamefont {Ionescu}, \citenamefont {Kisyov}, \citenamefont
  {Konki}, \citenamefont {Lazarus}, \citenamefont {Madurga}, \citenamefont
  {M\ifmmode~\u{a}\else \u{a}\fi{}rginean}, \citenamefont {M\ifmmode~\u{a}\else
  \u{a}\fi{}rginean}, \citenamefont {Mihai}, \citenamefont {Mihai},
  \citenamefont {Negret}, \citenamefont {Nowacki}, \citenamefont {Page},
  \citenamefont {Pakarinen}, \citenamefont {Pucknell}, \citenamefont {Rahkila},
  \citenamefont {Rapisarda}, \citenamefont {\ifmmode~\mbox{\c{S}}\else
  \c{S}\fi{}erban}, \citenamefont {Sotty}, \citenamefont {Stan}, \citenamefont
  {St\ifmmode~\u{a}\else \u{a}\fi{}noiu}, \citenamefont {Tengblad},
  \citenamefont {Turturic\ifmmode~\u{a}\else \u{a}\fi{}}, \citenamefont
  {Van~Duppen}, \citenamefont {Warr}, \citenamefont {Dessagne}, \citenamefont
  {Stora}, \citenamefont {Borcea}, \citenamefont {C\ifmmode~\u{a}\else
  \u{a}\fi{}linescu}, \citenamefont {Daugas}, \citenamefont {Filipescu},
  \citenamefont {Kuti}, \citenamefont {Franchoo}, \citenamefont {Gheorghe},
  \citenamefont {Morfouace}, \citenamefont {Morel}, \citenamefont {Mrazek},
  \citenamefont {Pietreanu}, \citenamefont {Sohler}, \citenamefont {Stefan},
  \citenamefont {\ifmmode \mbox{\c{S}}\else \c{S}\fi{}uv\ifmmode \u{a}\else
  \u{a}\fi{}il\ifmmode~\u{a}\else \u{a}\fi{}}, \citenamefont {Toma},\ and\
  \citenamefont {Ur}}]{Lica19}%
  \BibitemOpen
  \bibfield  {author} {\bibinfo {author} {\bibfnamefont {R.}~\bibnamefont
  {Lic\ifmmode~\u{a}\else \u{a}\fi{}}}, \bibinfo {author} {\bibfnamefont
  {F.}~\bibnamefont {Rotaru}}, \bibinfo {author} {\bibfnamefont {M.~J.~G.}\
  \bibnamefont {Borge}}, \bibinfo {author} {\bibfnamefont {S.}~\bibnamefont
  {Gr\'evy}}, \bibinfo {author} {\bibfnamefont {F.}~\bibnamefont {Negoi\ifmmode
  \mbox{\c{t}}\else \c{t}\fi{}\ifmmode~\u{a}\else \u{a}\fi{}}}, \bibinfo
  {author} {\bibfnamefont {A.}~\bibnamefont {Poves}}, \bibinfo {author}
  {\bibfnamefont {O.}~\bibnamefont {Sorlin}}, \bibinfo {author} {\bibfnamefont
  {A.~N.}\ \bibnamefont {Andreyev}}, \bibinfo {author} {\bibfnamefont
  {R.}~\bibnamefont {Borcea}}, \bibinfo {author} {\bibfnamefont
  {C.}~\bibnamefont {Costache}}, \bibinfo {author} {\bibfnamefont
  {H.}~\bibnamefont {De~Witte}}, \bibinfo {author} {\bibfnamefont {L.~M.}\
  \bibnamefont {Fraile}}, \bibinfo {author} {\bibfnamefont {P.~T.}\
  \bibnamefont {Greenlees}}, \bibinfo {author} {\bibfnamefont {M.}~\bibnamefont
  {Huyse}}, \bibinfo {author} {\bibfnamefont {A.}~\bibnamefont {Ionescu}},
  \bibinfo {author} {\bibfnamefont {S.}~\bibnamefont {Kisyov}}, \bibinfo
  {author} {\bibfnamefont {J.}~\bibnamefont {Konki}}, \bibinfo {author}
  {\bibfnamefont {I.}~\bibnamefont {Lazarus}}, \bibinfo {author} {\bibfnamefont
  {M.}~\bibnamefont {Madurga}}, \bibinfo {author} {\bibfnamefont
  {N.}~\bibnamefont {M\ifmmode~\u{a}\else \u{a}\fi{}rginean}}, \bibinfo
  {author} {\bibfnamefont {R.}~\bibnamefont {M\ifmmode~\u{a}\else
  \u{a}\fi{}rginean}}, \bibinfo {author} {\bibfnamefont {C.}~\bibnamefont
  {Mihai}}, \bibinfo {author} {\bibfnamefont {R.~E.}\ \bibnamefont {Mihai}},
  \bibinfo {author} {\bibfnamefont {A.}~\bibnamefont {Negret}}, \bibinfo
  {author} {\bibfnamefont {F.}~\bibnamefont {Nowacki}}, \bibinfo {author}
  {\bibfnamefont {R.~D.}\ \bibnamefont {Page}}, \bibinfo {author}
  {\bibfnamefont {J.}~\bibnamefont {Pakarinen}}, \bibinfo {author}
  {\bibfnamefont {V.}~\bibnamefont {Pucknell}}, \bibinfo {author}
  {\bibfnamefont {P.}~\bibnamefont {Rahkila}}, \bibinfo {author} {\bibfnamefont
  {E.}~\bibnamefont {Rapisarda}}, \bibinfo {author} {\bibfnamefont
  {A.}~\bibnamefont {\ifmmode~\mbox{\c{S}}\else \c{S}\fi{}erban}}, \bibinfo
  {author} {\bibfnamefont {C.~O.}\ \bibnamefont {Sotty}}, \bibinfo {author}
  {\bibfnamefont {L.}~\bibnamefont {Stan}}, \bibinfo {author} {\bibfnamefont
  {M.}~\bibnamefont {St\ifmmode~\u{a}\else \u{a}\fi{}noiu}}, \bibinfo {author}
  {\bibfnamefont {O.}~\bibnamefont {Tengblad}}, \bibinfo {author}
  {\bibfnamefont {A.}~\bibnamefont {Turturic\ifmmode~\u{a}\else \u{a}\fi{}}},
  \bibinfo {author} {\bibfnamefont {P.}~\bibnamefont {Van~Duppen}}, \bibinfo
  {author} {\bibfnamefont {N.}~\bibnamefont {Warr}}, \bibinfo {author}
  {\bibfnamefont {P.}~\bibnamefont {Dessagne}}, \bibinfo {author}
  {\bibfnamefont {T.}~\bibnamefont {Stora}}, \bibinfo {author} {\bibfnamefont
  {C.}~\bibnamefont {Borcea}}, \bibinfo {author} {\bibfnamefont
  {S.}~\bibnamefont {C\ifmmode~\u{a}\else \u{a}\fi{}linescu}}, \bibinfo
  {author} {\bibfnamefont {J.~M.}\ \bibnamefont {Daugas}}, \bibinfo {author}
  {\bibfnamefont {D.}~\bibnamefont {Filipescu}}, \bibinfo {author}
  {\bibfnamefont {I.}~\bibnamefont {Kuti}}, \bibinfo {author} {\bibfnamefont
  {S.}~\bibnamefont {Franchoo}}, \bibinfo {author} {\bibfnamefont
  {I.}~\bibnamefont {Gheorghe}}, \bibinfo {author} {\bibfnamefont
  {P.}~\bibnamefont {Morfouace}}, \bibinfo {author} {\bibfnamefont
  {P.}~\bibnamefont {Morel}}, \bibinfo {author} {\bibfnamefont
  {J.}~\bibnamefont {Mrazek}}, \bibinfo {author} {\bibfnamefont
  {D.}~\bibnamefont {Pietreanu}}, \bibinfo {author} {\bibfnamefont
  {D.}~\bibnamefont {Sohler}}, \bibinfo {author} {\bibfnamefont
  {I.}~\bibnamefont {Stefan}}, \bibinfo {author} {\bibfnamefont
  {R.}~\bibnamefont {\ifmmode \mbox{\c{S}}\else \c{S}\fi{}uv\ifmmode \u{a}\else
  \u{a}\fi{}il\ifmmode~\u{a}\else \u{a}\fi{}}}, \bibinfo {author}
  {\bibfnamefont {S.}~\bibnamefont {Toma}}, \ and\ \bibinfo {author}
  {\bibfnamefont {C.~A.}\ \bibnamefont {Ur}} (\bibinfo {collaboration} {IDS
  Collaboration}),\ }\href {\doibase 10.1103/PhysRevC.100.034306} {\bibfield
  {journal} {\bibinfo  {journal} {Phys. Rev. C}\ }\textbf {\bibinfo {volume}
  {100}},\ \bibinfo {pages} {034306} (\bibinfo {year} {2019})}\BibitemShut
  {NoStop}%
\bibitem [{\citenamefont {Ibbotson}\ \emph {et~al.}(1998)\citenamefont
  {Ibbotson}, \citenamefont {Glasmacher}, \citenamefont {Brown}, \citenamefont
  {Chen}, \citenamefont {Chromik}, \citenamefont {Cottle}, \citenamefont
  {Fauerbach}, \citenamefont {Kemper}, \citenamefont {Morrissey}, \citenamefont
  {Scheit},\ and\ \citenamefont {Thoennessen}}]{Ibbo98}%
  \BibitemOpen
  \bibfield  {author} {\bibinfo {author} {\bibfnamefont {R.~W.}\ \bibnamefont
  {Ibbotson}}, \bibinfo {author} {\bibfnamefont {T.}~\bibnamefont
  {Glasmacher}}, \bibinfo {author} {\bibfnamefont {B.~A.}\ \bibnamefont
  {Brown}}, \bibinfo {author} {\bibfnamefont {L.}~\bibnamefont {Chen}},
  \bibinfo {author} {\bibfnamefont {M.~J.}\ \bibnamefont {Chromik}}, \bibinfo
  {author} {\bibfnamefont {P.~D.}\ \bibnamefont {Cottle}}, \bibinfo {author}
  {\bibfnamefont {M.}~\bibnamefont {Fauerbach}}, \bibinfo {author}
  {\bibfnamefont {K.~W.}\ \bibnamefont {Kemper}}, \bibinfo {author}
  {\bibfnamefont {D.~J.}\ \bibnamefont {Morrissey}}, \bibinfo {author}
  {\bibfnamefont {H.}~\bibnamefont {Scheit}}, \ and\ \bibinfo {author}
  {\bibfnamefont {M.}~\bibnamefont {Thoennessen}},\ }\href {\doibase
  10.1103/PhysRevLett.80.2081} {\bibfield  {journal} {\bibinfo  {journal}
  {Phys. Rev. Lett.}\ }\textbf {\bibinfo {volume} {80}},\ \bibinfo {pages}
  {2081} (\bibinfo {year} {1998})}\BibitemShut {NoStop}%
\bibitem [{\citenamefont {Ascher}\ \emph {et~al.}(2019)\citenamefont {Ascher},
  \citenamefont {Althubiti}, \citenamefont {Atanasov}, \citenamefont {Blaum},
  \citenamefont {Cakirli}, \citenamefont {Gr\'evy}, \citenamefont {Herfurth},
  \citenamefont {Kreim}, \citenamefont {Lunney}, \citenamefont {Manea},
  \citenamefont {Neidherr}, \citenamefont {Rosenbusch}, \citenamefont
  {Schweikhard}, \citenamefont {Welker}, \citenamefont {Wienholtz},
  \citenamefont {Wolf},\ and\ \citenamefont {Zuber}}]{Asch19}%
  \BibitemOpen
  \bibfield  {author} {\bibinfo {author} {\bibfnamefont {P.}~\bibnamefont
  {Ascher}}, \bibinfo {author} {\bibfnamefont {N.}~\bibnamefont {Althubiti}},
  \bibinfo {author} {\bibfnamefont {D.}~\bibnamefont {Atanasov}}, \bibinfo
  {author} {\bibfnamefont {K.}~\bibnamefont {Blaum}}, \bibinfo {author}
  {\bibfnamefont {R.~B.}\ \bibnamefont {Cakirli}}, \bibinfo {author}
  {\bibfnamefont {S.}~\bibnamefont {Gr\'evy}}, \bibinfo {author} {\bibfnamefont
  {F.}~\bibnamefont {Herfurth}}, \bibinfo {author} {\bibfnamefont
  {S.}~\bibnamefont {Kreim}}, \bibinfo {author} {\bibfnamefont
  {D.}~\bibnamefont {Lunney}}, \bibinfo {author} {\bibfnamefont
  {V.}~\bibnamefont {Manea}}, \bibinfo {author} {\bibfnamefont
  {D.}~\bibnamefont {Neidherr}}, \bibinfo {author} {\bibfnamefont
  {M.}~\bibnamefont {Rosenbusch}}, \bibinfo {author} {\bibfnamefont
  {L.}~\bibnamefont {Schweikhard}}, \bibinfo {author} {\bibfnamefont
  {A.}~\bibnamefont {Welker}}, \bibinfo {author} {\bibfnamefont
  {F.}~\bibnamefont {Wienholtz}}, \bibinfo {author} {\bibfnamefont {R.~N.}\
  \bibnamefont {Wolf}}, \ and\ \bibinfo {author} {\bibfnamefont
  {K.}~\bibnamefont {Zuber}},\ }\href {\doibase 10.1103/PhysRevC.100.014304}
  {\bibfield  {journal} {\bibinfo  {journal} {Phys. Rev. C}\ }\textbf {\bibinfo
  {volume} {100}},\ \bibinfo {pages} {014304} (\bibinfo {year}
  {2019})}\BibitemShut {NoStop}%
\bibitem [{\citenamefont {Lic\ifmmode~\u{a}\else \u{a}\fi{}}\ \emph
  {et~al.}(2017)\citenamefont {Lic\ifmmode~\u{a}\else \u{a}\fi{}},
  \citenamefont {Rotaru}, \citenamefont {Borge}, \citenamefont {Gr\'evy},
  \citenamefont {Negoi\ifmmode \mbox{\c{t}}\else \c{t}\fi{}\ifmmode~\u{a}\else
  \u{a}\fi{}}, \citenamefont {Poves}, \citenamefont {Sorlin}, \citenamefont
  {Andreyev}, \citenamefont {Borcea}, \citenamefont {Costache}, \citenamefont
  {De~Witte}, \citenamefont {Fraile}, \citenamefont {Greenlees}, \citenamefont
  {Huyse}, \citenamefont {Ionescu}, \citenamefont {Kisyov}, \citenamefont
  {Konki}, \citenamefont {Lazarus}, \citenamefont {Madurga}, \citenamefont
  {M\ifmmode~\u{a}\else \u{a}\fi{}rginean}, \citenamefont {M\ifmmode~\u{a}\else
  \u{a}\fi{}rginean}, \citenamefont {Mihai}, \citenamefont {Mihai},
  \citenamefont {Negret}, \citenamefont {Page}, \citenamefont {Pakarinen},
  \citenamefont {Pascu}, \citenamefont {Pucknell}, \citenamefont {Rahkila},
  \citenamefont {Rapisarda}, \citenamefont {\ifmmode~\mbox{\c{S}}\else
  \c{S}\fi{}erban}, \citenamefont {Sotty}, \citenamefont {Stan}, \citenamefont
  {St\ifmmode~\u{a}\else \u{a}\fi{}noiu}, \citenamefont {Tengblad},
  \citenamefont {Turturic\ifmmode~\u{a}\else \u{a}\fi{}}, \citenamefont
  {Van~Duppen}, \citenamefont {Wadsworth},\ and\ \citenamefont
  {Warr}}]{Lica17}%
  \BibitemOpen
  \bibfield  {author} {\bibinfo {author} {\bibfnamefont {R.}~\bibnamefont
  {Lic\ifmmode~\u{a}\else \u{a}\fi{}}}, \bibinfo {author} {\bibfnamefont
  {F.}~\bibnamefont {Rotaru}}, \bibinfo {author} {\bibfnamefont {M.~J.~G.}\
  \bibnamefont {Borge}}, \bibinfo {author} {\bibfnamefont {S.}~\bibnamefont
  {Gr\'evy}}, \bibinfo {author} {\bibfnamefont {F.}~\bibnamefont {Negoi\ifmmode
  \mbox{\c{t}}\else \c{t}\fi{}\ifmmode~\u{a}\else \u{a}\fi{}}}, \bibinfo
  {author} {\bibfnamefont {A.}~\bibnamefont {Poves}}, \bibinfo {author}
  {\bibfnamefont {O.}~\bibnamefont {Sorlin}}, \bibinfo {author} {\bibfnamefont
  {A.~N.}\ \bibnamefont {Andreyev}}, \bibinfo {author} {\bibfnamefont
  {R.}~\bibnamefont {Borcea}}, \bibinfo {author} {\bibfnamefont
  {C.}~\bibnamefont {Costache}}, \bibinfo {author} {\bibfnamefont
  {H.}~\bibnamefont {De~Witte}}, \bibinfo {author} {\bibfnamefont {L.~M.}\
  \bibnamefont {Fraile}}, \bibinfo {author} {\bibfnamefont {P.~T.}\
  \bibnamefont {Greenlees}}, \bibinfo {author} {\bibfnamefont {M.}~\bibnamefont
  {Huyse}}, \bibinfo {author} {\bibfnamefont {A.}~\bibnamefont {Ionescu}},
  \bibinfo {author} {\bibfnamefont {S.}~\bibnamefont {Kisyov}}, \bibinfo
  {author} {\bibfnamefont {J.}~\bibnamefont {Konki}}, \bibinfo {author}
  {\bibfnamefont {I.}~\bibnamefont {Lazarus}}, \bibinfo {author} {\bibfnamefont
  {M.}~\bibnamefont {Madurga}}, \bibinfo {author} {\bibfnamefont
  {N.}~\bibnamefont {M\ifmmode~\u{a}\else \u{a}\fi{}rginean}}, \bibinfo
  {author} {\bibfnamefont {R.}~\bibnamefont {M\ifmmode~\u{a}\else
  \u{a}\fi{}rginean}}, \bibinfo {author} {\bibfnamefont {C.}~\bibnamefont
  {Mihai}}, \bibinfo {author} {\bibfnamefont {R.~E.}\ \bibnamefont {Mihai}},
  \bibinfo {author} {\bibfnamefont {A.}~\bibnamefont {Negret}}, \bibinfo
  {author} {\bibfnamefont {R.~D.}\ \bibnamefont {Page}}, \bibinfo {author}
  {\bibfnamefont {J.}~\bibnamefont {Pakarinen}}, \bibinfo {author}
  {\bibfnamefont {S.}~\bibnamefont {Pascu}}, \bibinfo {author} {\bibfnamefont
  {V.}~\bibnamefont {Pucknell}}, \bibinfo {author} {\bibfnamefont
  {P.}~\bibnamefont {Rahkila}}, \bibinfo {author} {\bibfnamefont
  {E.}~\bibnamefont {Rapisarda}}, \bibinfo {author} {\bibfnamefont
  {A.}~\bibnamefont {\ifmmode~\mbox{\c{S}}\else \c{S}\fi{}erban}}, \bibinfo
  {author} {\bibfnamefont {C.~O.}\ \bibnamefont {Sotty}}, \bibinfo {author}
  {\bibfnamefont {L.}~\bibnamefont {Stan}}, \bibinfo {author} {\bibfnamefont
  {M.}~\bibnamefont {St\ifmmode~\u{a}\else \u{a}\fi{}noiu}}, \bibinfo {author}
  {\bibfnamefont {O.}~\bibnamefont {Tengblad}}, \bibinfo {author}
  {\bibfnamefont {A.}~\bibnamefont {Turturic\ifmmode~\u{a}\else \u{a}\fi{}}},
  \bibinfo {author} {\bibfnamefont {P.}~\bibnamefont {Van~Duppen}}, \bibinfo
  {author} {\bibfnamefont {R.}~\bibnamefont {Wadsworth}}, \ and\ \bibinfo
  {author} {\bibfnamefont {N.}~\bibnamefont {Warr}} (\bibinfo {collaboration}
  {IDS Collaboration}),\ }\href {\doibase 10.1103/PhysRevC.95.021301}
  {\bibfield  {journal} {\bibinfo  {journal} {Phys. Rev. C}\ }\textbf {\bibinfo
  {volume} {95}},\ \bibinfo {pages} {021301} (\bibinfo {year}
  {2017})}\BibitemShut {NoStop}%
\bibitem [{\citenamefont {Rotaru}\ \emph {et~al.}(2012)\citenamefont {Rotaru},
  \citenamefont {Negoita}, \citenamefont {Gr\'evy}, \citenamefont {Mrazek},
  \citenamefont {Lukyanov}, \citenamefont {Nowacki}, \citenamefont {Poves},
  \citenamefont {Sorlin}, \citenamefont {Borcea}, \citenamefont {Borcea},
  \citenamefont {Buta}, \citenamefont {C\'aceres}, \citenamefont {Calinescu},
  \citenamefont {Chevrier}, \citenamefont {Dombr\'adi}, \citenamefont {Daugas},
  \citenamefont {Lebhertz}, \citenamefont {Penionzhkevich}, \citenamefont
  {Petrone}, \citenamefont {Sohler}, \citenamefont {Stanoiu},\ and\
  \citenamefont {Thomas}}]{Rota12}%
  \BibitemOpen
  \bibfield  {author} {\bibinfo {author} {\bibfnamefont {F.}~\bibnamefont
  {Rotaru}}, \bibinfo {author} {\bibfnamefont {F.}~\bibnamefont {Negoita}},
  \bibinfo {author} {\bibfnamefont {S.}~\bibnamefont {Gr\'evy}}, \bibinfo
  {author} {\bibfnamefont {J.}~\bibnamefont {Mrazek}}, \bibinfo {author}
  {\bibfnamefont {S.}~\bibnamefont {Lukyanov}}, \bibinfo {author}
  {\bibfnamefont {F.}~\bibnamefont {Nowacki}}, \bibinfo {author} {\bibfnamefont
  {A.}~\bibnamefont {Poves}}, \bibinfo {author} {\bibfnamefont
  {O.}~\bibnamefont {Sorlin}}, \bibinfo {author} {\bibfnamefont
  {C.}~\bibnamefont {Borcea}}, \bibinfo {author} {\bibfnamefont
  {R.}~\bibnamefont {Borcea}}, \bibinfo {author} {\bibfnamefont
  {A.}~\bibnamefont {Buta}}, \bibinfo {author} {\bibfnamefont {L.}~\bibnamefont
  {C\'aceres}}, \bibinfo {author} {\bibfnamefont {S.}~\bibnamefont
  {Calinescu}}, \bibinfo {author} {\bibfnamefont {R.}~\bibnamefont {Chevrier}},
  \bibinfo {author} {\bibfnamefont {Z.}~\bibnamefont {Dombr\'adi}}, \bibinfo
  {author} {\bibfnamefont {J.~M.}\ \bibnamefont {Daugas}}, \bibinfo {author}
  {\bibfnamefont {D.}~\bibnamefont {Lebhertz}}, \bibinfo {author}
  {\bibfnamefont {Y.}~\bibnamefont {Penionzhkevich}}, \bibinfo {author}
  {\bibfnamefont {C.}~\bibnamefont {Petrone}}, \bibinfo {author} {\bibfnamefont
  {D.}~\bibnamefont {Sohler}}, \bibinfo {author} {\bibfnamefont
  {M.}~\bibnamefont {Stanoiu}}, \ and\ \bibinfo {author} {\bibfnamefont
  {J.~C.}\ \bibnamefont {Thomas}},\ }\href {\doibase
  10.1103/PhysRevLett.109.092503} {\bibfield  {journal} {\bibinfo  {journal}
  {Phys. Rev. Lett.}\ }\textbf {\bibinfo {volume} {109}},\ \bibinfo {pages}
  {092503} (\bibinfo {year} {2012})}\BibitemShut {NoStop}%
\bibitem [{\citenamefont {Han}\ \emph {et~al.}(2017)\citenamefont {Han},
  \citenamefont {Li}, \citenamefont {Jiang}, \citenamefont {Li}, \citenamefont
  {Hua}, \citenamefont {Zhang}, \citenamefont {Yuan}, \citenamefont {Jiang},
  \citenamefont {Ye}, \citenamefont {Li}, \citenamefont {Li}, \citenamefont
  {Xu}, \citenamefont {Chen}, \citenamefont {Meng}, \citenamefont {Wang},
  \citenamefont {Xu}, \citenamefont {Sun}, \citenamefont {Wang}, \citenamefont
  {Wu}, \citenamefont {Niu}, \citenamefont {Li}, \citenamefont {He},
  \citenamefont {Jiang}, \citenamefont {Li}, \citenamefont {Zang},
  \citenamefont {Feng}, \citenamefont {Chen}, \citenamefont {Liu},
  \citenamefont {Chen}, \citenamefont {Xu}, \citenamefont {Hu}, \citenamefont
  {Yang}, \citenamefont {Ma}, \citenamefont {Ma}, \citenamefont {Jin},
  \citenamefont {Bai}, \citenamefont {Huang}, \citenamefont {Zhou},
  \citenamefont {Ma}, \citenamefont {Li}, \citenamefont {Zhou}, \citenamefont
  {Zhang}, \citenamefont {Xiao},\ and\ \citenamefont {Zhan}}]{Han17}%
  \BibitemOpen
  \bibfield  {author} {\bibinfo {author} {\bibfnamefont {R.}~\bibnamefont
  {Han}}, \bibinfo {author} {\bibfnamefont {X.}~\bibnamefont {Li}}, \bibinfo
  {author} {\bibfnamefont {W.}~\bibnamefont {Jiang}}, \bibinfo {author}
  {\bibfnamefont {Z.}~\bibnamefont {Li}}, \bibinfo {author} {\bibfnamefont
  {H.}~\bibnamefont {Hua}}, \bibinfo {author} {\bibfnamefont {S.}~\bibnamefont
  {Zhang}}, \bibinfo {author} {\bibfnamefont {C.}~\bibnamefont {Yuan}},
  \bibinfo {author} {\bibfnamefont {D.}~\bibnamefont {Jiang}}, \bibinfo
  {author} {\bibfnamefont {Y.}~\bibnamefont {Ye}}, \bibinfo {author}
  {\bibfnamefont {J.}~\bibnamefont {Li}}, \bibinfo {author} {\bibfnamefont
  {Z.}~\bibnamefont {Li}}, \bibinfo {author} {\bibfnamefont {F.}~\bibnamefont
  {Xu}}, \bibinfo {author} {\bibfnamefont {Q.}~\bibnamefont {Chen}}, \bibinfo
  {author} {\bibfnamefont {J.}~\bibnamefont {Meng}}, \bibinfo {author}
  {\bibfnamefont {J.}~\bibnamefont {Wang}}, \bibinfo {author} {\bibfnamefont
  {C.}~\bibnamefont {Xu}}, \bibinfo {author} {\bibfnamefont {Y.}~\bibnamefont
  {Sun}}, \bibinfo {author} {\bibfnamefont {C.}~\bibnamefont {Wang}}, \bibinfo
  {author} {\bibfnamefont {H.}~\bibnamefont {Wu}}, \bibinfo {author}
  {\bibfnamefont {C.}~\bibnamefont {Niu}}, \bibinfo {author} {\bibfnamefont
  {C.}~\bibnamefont {Li}}, \bibinfo {author} {\bibfnamefont {C.}~\bibnamefont
  {He}}, \bibinfo {author} {\bibfnamefont {W.}~\bibnamefont {Jiang}}, \bibinfo
  {author} {\bibfnamefont {P.}~\bibnamefont {Li}}, \bibinfo {author}
  {\bibfnamefont {H.}~\bibnamefont {Zang}}, \bibinfo {author} {\bibfnamefont
  {J.}~\bibnamefont {Feng}}, \bibinfo {author} {\bibfnamefont {S.}~\bibnamefont
  {Chen}}, \bibinfo {author} {\bibfnamefont {Q.}~\bibnamefont {Liu}}, \bibinfo
  {author} {\bibfnamefont {X.}~\bibnamefont {Chen}}, \bibinfo {author}
  {\bibfnamefont {H.}~\bibnamefont {Xu}}, \bibinfo {author} {\bibfnamefont
  {Z.}~\bibnamefont {Hu}}, \bibinfo {author} {\bibfnamefont {Y.}~\bibnamefont
  {Yang}}, \bibinfo {author} {\bibfnamefont {P.}~\bibnamefont {Ma}}, \bibinfo
  {author} {\bibfnamefont {J.}~\bibnamefont {Ma}}, \bibinfo {author}
  {\bibfnamefont {S.}~\bibnamefont {Jin}}, \bibinfo {author} {\bibfnamefont
  {Z.}~\bibnamefont {Bai}}, \bibinfo {author} {\bibfnamefont {M.}~\bibnamefont
  {Huang}}, \bibinfo {author} {\bibfnamefont {Y.}~\bibnamefont {Zhou}},
  \bibinfo {author} {\bibfnamefont {W.}~\bibnamefont {Ma}}, \bibinfo {author}
  {\bibfnamefont {Y.}~\bibnamefont {Li}}, \bibinfo {author} {\bibfnamefont
  {X.}~\bibnamefont {Zhou}}, \bibinfo {author} {\bibfnamefont {Y.}~\bibnamefont
  {Zhang}}, \bibinfo {author} {\bibfnamefont {G.}~\bibnamefont {Xiao}}, \ and\
  \bibinfo {author} {\bibfnamefont {W.}~\bibnamefont {Zhan}},\ }\href {\doibase
  https://doi.org/10.1016/j.physletb.2017.07.007} {\bibfield  {journal}
  {\bibinfo  {journal} {Physics Letters B}\ }\textbf {\bibinfo {volume}
  {772}},\ \bibinfo {pages} {529 } (\bibinfo {year} {2017})}\BibitemShut
  {NoStop}%
\bibitem [{\citenamefont {Paschalis}\ \emph {et~al.}(2011)\citenamefont
  {Paschalis}, \citenamefont {Fallon}, \citenamefont {Macchiavelli},
  \citenamefont {Petri}, \citenamefont {Bender}, \citenamefont {Carpenter},
  \citenamefont {Chen}, \citenamefont {Chiara}, \citenamefont {Clark},
  \citenamefont {Cromaz}, \citenamefont {Gros}, \citenamefont {Hamilton},
  \citenamefont {Hoffman}, \citenamefont {Janssens}, \citenamefont {Lauritsen},
  \citenamefont {Lee}, \citenamefont {Lister}, \citenamefont {McCutchan},
  \citenamefont {Phair}, \citenamefont {Reviol}, \citenamefont {Sarantites},
  \citenamefont {Seweryniak}, \citenamefont {Tabor}, \citenamefont {Toh},
  \citenamefont {Wiedeking},\ and\ \citenamefont {Zhu}}]{Pasch11}%
  \BibitemOpen
  \bibfield  {author} {\bibinfo {author} {\bibfnamefont {S.}~\bibnamefont
  {Paschalis}}, \bibinfo {author} {\bibfnamefont {P.}~\bibnamefont {Fallon}},
  \bibinfo {author} {\bibfnamefont {A.~O.}\ \bibnamefont {Macchiavelli}},
  \bibinfo {author} {\bibfnamefont {M.}~\bibnamefont {Petri}}, \bibinfo
  {author} {\bibfnamefont {P.~C.}\ \bibnamefont {Bender}}, \bibinfo {author}
  {\bibfnamefont {M.~P.}\ \bibnamefont {Carpenter}}, \bibinfo {author}
  {\bibfnamefont {X.}~\bibnamefont {Chen}}, \bibinfo {author} {\bibfnamefont
  {C.~J.}\ \bibnamefont {Chiara}}, \bibinfo {author} {\bibfnamefont {R.~M.}\
  \bibnamefont {Clark}}, \bibinfo {author} {\bibfnamefont {M.}~\bibnamefont
  {Cromaz}}, \bibinfo {author} {\bibfnamefont {S.}~\bibnamefont {Gros}},
  \bibinfo {author} {\bibfnamefont {L.}~\bibnamefont {Hamilton}}, \bibinfo
  {author} {\bibfnamefont {C.~R.}\ \bibnamefont {Hoffman}}, \bibinfo {author}
  {\bibfnamefont {R.~V.~F.}\ \bibnamefont {Janssens}}, \bibinfo {author}
  {\bibfnamefont {T.}~\bibnamefont {Lauritsen}}, \bibinfo {author}
  {\bibfnamefont {I.~Y.}\ \bibnamefont {Lee}}, \bibinfo {author} {\bibfnamefont
  {C.~J.}\ \bibnamefont {Lister}}, \bibinfo {author} {\bibfnamefont {E.~A.}\
  \bibnamefont {McCutchan}}, \bibinfo {author} {\bibfnamefont {L.}~\bibnamefont
  {Phair}}, \bibinfo {author} {\bibfnamefont {W.}~\bibnamefont {Reviol}},
  \bibinfo {author} {\bibfnamefont {D.~G.}\ \bibnamefont {Sarantites}},
  \bibinfo {author} {\bibfnamefont {D.}~\bibnamefont {Seweryniak}}, \bibinfo
  {author} {\bibfnamefont {S.~L.}\ \bibnamefont {Tabor}}, \bibinfo {author}
  {\bibfnamefont {Y.}~\bibnamefont {Toh}}, \bibinfo {author} {\bibfnamefont
  {M.}~\bibnamefont {Wiedeking}}, \ and\ \bibinfo {author} {\bibfnamefont
  {S.}~\bibnamefont {Zhu}},\ }\href {\doibase 10.1088/1742-6596/312/9/092050}
  {\bibfield  {journal} {\bibinfo  {journal} {Journal of Physics: Conference
  Series}\ }\textbf {\bibinfo {volume} {312}},\ \bibinfo {pages} {092050}
  (\bibinfo {year} {2011})}\BibitemShut {NoStop}%
\bibitem [{\citenamefont {Wiedeking}\ \emph {et~al.}(2008)\citenamefont
  {Wiedeking}, \citenamefont {Rodriguez-Vieitez}, \citenamefont {Fallon},
  \citenamefont {Carpenter}, \citenamefont {Clark}, \citenamefont {Cline},
  \citenamefont {Cromaz}, \citenamefont {Descovich}, \citenamefont {Janssens},
  \citenamefont {Lee}, \citenamefont {Deleplanque}, \citenamefont
  {Macchiavelli}, \citenamefont {Stephens}, \citenamefont {Teng}, \citenamefont
  {Wang}, \citenamefont {Ward}, \citenamefont {Wu}, \citenamefont {Zhu},
  \citenamefont {Otsuka}, \citenamefont {Utsuno},\ and\ \citenamefont
  {Volya}}]{Wied08}%
  \BibitemOpen
  \bibfield  {author} {\bibinfo {author} {\bibfnamefont {M.}~\bibnamefont
  {Wiedeking}}, \bibinfo {author} {\bibfnamefont {E.}~\bibnamefont
  {Rodriguez-Vieitez}}, \bibinfo {author} {\bibfnamefont {P.}~\bibnamefont
  {Fallon}}, \bibinfo {author} {\bibfnamefont {M.~P.}\ \bibnamefont
  {Carpenter}}, \bibinfo {author} {\bibfnamefont {R.~M.}\ \bibnamefont
  {Clark}}, \bibinfo {author} {\bibfnamefont {D.}~\bibnamefont {Cline}},
  \bibinfo {author} {\bibfnamefont {M.}~\bibnamefont {Cromaz}}, \bibinfo
  {author} {\bibfnamefont {M.}~\bibnamefont {Descovich}}, \bibinfo {author}
  {\bibfnamefont {R.~V.~F.}\ \bibnamefont {Janssens}}, \bibinfo {author}
  {\bibfnamefont {I.-Y.}\ \bibnamefont {Lee}}, \bibinfo {author} {\bibfnamefont
  {M.-A.}\ \bibnamefont {Deleplanque}}, \bibinfo {author} {\bibfnamefont
  {A.~O.}\ \bibnamefont {Macchiavelli}}, \bibinfo {author} {\bibfnamefont
  {F.~S.}\ \bibnamefont {Stephens}}, \bibinfo {author} {\bibfnamefont
  {R.}~\bibnamefont {Teng}}, \bibinfo {author} {\bibfnamefont {X.}~\bibnamefont
  {Wang}}, \bibinfo {author} {\bibfnamefont {D.}~\bibnamefont {Ward}}, \bibinfo
  {author} {\bibfnamefont {C.~Y.}\ \bibnamefont {Wu}}, \bibinfo {author}
  {\bibfnamefont {S.}~\bibnamefont {Zhu}}, \bibinfo {author} {\bibfnamefont
  {T.}~\bibnamefont {Otsuka}}, \bibinfo {author} {\bibfnamefont
  {Y.}~\bibnamefont {Utsuno}}, \ and\ \bibinfo {author} {\bibfnamefont
  {A.}~\bibnamefont {Volya}},\ }\href {\doibase 10.1103/PhysRevC.78.037302}
  {\bibfield  {journal} {\bibinfo  {journal} {Phys. Rev. C}\ }\textbf {\bibinfo
  {volume} {78}},\ \bibinfo {pages} {037302} (\bibinfo {year}
  {2008})}\BibitemShut {NoStop}%
\bibitem [{\citenamefont {Wimmer}\ \emph {et~al.}(2010)\citenamefont {Wimmer},
  \citenamefont {Kr\"oll}, \citenamefont {Kr\"ucken}, \citenamefont
  {Bildstein}, \citenamefont {Gernh\"auser}, \citenamefont {Bastin},
  \citenamefont {Bree}, \citenamefont {Diriken}, \citenamefont {Van~Duppen},
  \citenamefont {Huyse}, \citenamefont {Patronis}, \citenamefont {Vermaelen},
  \citenamefont {Voulot}, \citenamefont {Van~de Walle}, \citenamefont
  {Wenander}, \citenamefont {Fraile}, \citenamefont {Chapman}, \citenamefont
  {Hadinia}, \citenamefont {Orlandi}, \citenamefont {Smith}, \citenamefont
  {Lutter}, \citenamefont {Thirolf}, \citenamefont {Labiche}, \citenamefont
  {Blazhev}, \citenamefont {Kalk\"uhler}, \citenamefont {Reiter}, \citenamefont
  {Seidlitz}, \citenamefont {Warr}, \citenamefont {Macchiavelli}, \citenamefont
  {Jeppesen}, \citenamefont {Fiori}, \citenamefont {Georgiev}, \citenamefont
  {Schrieder}, \citenamefont {Das~Gupta}, \citenamefont {Lo~Bianco},
  \citenamefont {Nardelli}, \citenamefont {Butterworth}, \citenamefont
  {Johansen},\ and\ \citenamefont {Riisager}}]{Wimm10}%
  \BibitemOpen
  \bibfield  {author} {\bibinfo {author} {\bibfnamefont {K.}~\bibnamefont
  {Wimmer}}, \bibinfo {author} {\bibfnamefont {T.}~\bibnamefont {Kr\"oll}},
  \bibinfo {author} {\bibfnamefont {R.}~\bibnamefont {Kr\"ucken}}, \bibinfo
  {author} {\bibfnamefont {V.}~\bibnamefont {Bildstein}}, \bibinfo {author}
  {\bibfnamefont {R.}~\bibnamefont {Gernh\"auser}}, \bibinfo {author}
  {\bibfnamefont {B.}~\bibnamefont {Bastin}}, \bibinfo {author} {\bibfnamefont
  {N.}~\bibnamefont {Bree}}, \bibinfo {author} {\bibfnamefont {J.}~\bibnamefont
  {Diriken}}, \bibinfo {author} {\bibfnamefont {P.}~\bibnamefont {Van~Duppen}},
  \bibinfo {author} {\bibfnamefont {M.}~\bibnamefont {Huyse}}, \bibinfo
  {author} {\bibfnamefont {N.}~\bibnamefont {Patronis}}, \bibinfo {author}
  {\bibfnamefont {P.}~\bibnamefont {Vermaelen}}, \bibinfo {author}
  {\bibfnamefont {D.}~\bibnamefont {Voulot}}, \bibinfo {author} {\bibfnamefont
  {J.}~\bibnamefont {Van~de Walle}}, \bibinfo {author} {\bibfnamefont
  {F.}~\bibnamefont {Wenander}}, \bibinfo {author} {\bibfnamefont {L.~M.}\
  \bibnamefont {Fraile}}, \bibinfo {author} {\bibfnamefont {R.}~\bibnamefont
  {Chapman}}, \bibinfo {author} {\bibfnamefont {B.}~\bibnamefont {Hadinia}},
  \bibinfo {author} {\bibfnamefont {R.}~\bibnamefont {Orlandi}}, \bibinfo
  {author} {\bibfnamefont {J.~F.}\ \bibnamefont {Smith}}, \bibinfo {author}
  {\bibfnamefont {R.}~\bibnamefont {Lutter}}, \bibinfo {author} {\bibfnamefont
  {P.~G.}\ \bibnamefont {Thirolf}}, \bibinfo {author} {\bibfnamefont
  {M.}~\bibnamefont {Labiche}}, \bibinfo {author} {\bibfnamefont
  {A.}~\bibnamefont {Blazhev}}, \bibinfo {author} {\bibfnamefont
  {M.}~\bibnamefont {Kalk\"uhler}}, \bibinfo {author} {\bibfnamefont
  {P.}~\bibnamefont {Reiter}}, \bibinfo {author} {\bibfnamefont
  {M.}~\bibnamefont {Seidlitz}}, \bibinfo {author} {\bibfnamefont
  {N.}~\bibnamefont {Warr}}, \bibinfo {author} {\bibfnamefont {A.~O.}\
  \bibnamefont {Macchiavelli}}, \bibinfo {author} {\bibfnamefont {H.~B.}\
  \bibnamefont {Jeppesen}}, \bibinfo {author} {\bibfnamefont {E.}~\bibnamefont
  {Fiori}}, \bibinfo {author} {\bibfnamefont {G.}~\bibnamefont {Georgiev}},
  \bibinfo {author} {\bibfnamefont {G.}~\bibnamefont {Schrieder}}, \bibinfo
  {author} {\bibfnamefont {S.}~\bibnamefont {Das~Gupta}}, \bibinfo {author}
  {\bibfnamefont {G.}~\bibnamefont {Lo~Bianco}}, \bibinfo {author}
  {\bibfnamefont {S.}~\bibnamefont {Nardelli}}, \bibinfo {author}
  {\bibfnamefont {J.}~\bibnamefont {Butterworth}}, \bibinfo {author}
  {\bibfnamefont {J.}~\bibnamefont {Johansen}}, \ and\ \bibinfo {author}
  {\bibfnamefont {K.}~\bibnamefont {Riisager}},\ }\href {\doibase
  10.1103/PhysRevLett.105.252501} {\bibfield  {journal} {\bibinfo  {journal}
  {Phys. Rev. Lett.}\ }\textbf {\bibinfo {volume} {105}},\ \bibinfo {pages}
  {252501} (\bibinfo {year} {2010})}\BibitemShut {NoStop}%
\bibitem [{\citenamefont {Crawford}\ \emph {et~al.}(2016)\citenamefont
  {Crawford}, \citenamefont {Fallon}, \citenamefont {Macchiavelli},
  \citenamefont {Poves}, \citenamefont {Bader}, \citenamefont {Bazin},
  \citenamefont {Bowry}, \citenamefont {Campbell}, \citenamefont {Carpenter},
  \citenamefont {Clark}, \citenamefont {Cromaz}, \citenamefont {Gade},
  \citenamefont {Ideguchi}, \citenamefont {Iwasaki}, \citenamefont {Langer},
  \citenamefont {Lee}, \citenamefont {Loelius}, \citenamefont {Lunderberg},
  \citenamefont {Morse}, \citenamefont {Richard}, \citenamefont {Rissanen},
  \citenamefont {Smalley}, \citenamefont {Stroberg}, \citenamefont {Weisshaar},
  \citenamefont {Whitmore}, \citenamefont {Wiens}, \citenamefont {Williams},
  \citenamefont {Wimmer},\ and\ \citenamefont {Yamamato}}]{Craw16}%
  \BibitemOpen
  \bibfield  {author} {\bibinfo {author} {\bibfnamefont {H.~L.}\ \bibnamefont
  {Crawford}}, \bibinfo {author} {\bibfnamefont {P.}~\bibnamefont {Fallon}},
  \bibinfo {author} {\bibfnamefont {A.~O.}\ \bibnamefont {Macchiavelli}},
  \bibinfo {author} {\bibfnamefont {A.}~\bibnamefont {Poves}}, \bibinfo
  {author} {\bibfnamefont {V.~M.}\ \bibnamefont {Bader}}, \bibinfo {author}
  {\bibfnamefont {D.}~\bibnamefont {Bazin}}, \bibinfo {author} {\bibfnamefont
  {M.}~\bibnamefont {Bowry}}, \bibinfo {author} {\bibfnamefont {C.~M.}\
  \bibnamefont {Campbell}}, \bibinfo {author} {\bibfnamefont {M.~P.}\
  \bibnamefont {Carpenter}}, \bibinfo {author} {\bibfnamefont {R.~M.}\
  \bibnamefont {Clark}}, \bibinfo {author} {\bibfnamefont {M.}~\bibnamefont
  {Cromaz}}, \bibinfo {author} {\bibfnamefont {A.}~\bibnamefont {Gade}},
  \bibinfo {author} {\bibfnamefont {E.}~\bibnamefont {Ideguchi}}, \bibinfo
  {author} {\bibfnamefont {H.}~\bibnamefont {Iwasaki}}, \bibinfo {author}
  {\bibfnamefont {C.}~\bibnamefont {Langer}}, \bibinfo {author} {\bibfnamefont
  {I.~Y.}\ \bibnamefont {Lee}}, \bibinfo {author} {\bibfnamefont
  {C.}~\bibnamefont {Loelius}}, \bibinfo {author} {\bibfnamefont
  {E.}~\bibnamefont {Lunderberg}}, \bibinfo {author} {\bibfnamefont
  {C.}~\bibnamefont {Morse}}, \bibinfo {author} {\bibfnamefont {A.~L.}\
  \bibnamefont {Richard}}, \bibinfo {author} {\bibfnamefont {J.}~\bibnamefont
  {Rissanen}}, \bibinfo {author} {\bibfnamefont {D.}~\bibnamefont {Smalley}},
  \bibinfo {author} {\bibfnamefont {S.~R.}\ \bibnamefont {Stroberg}}, \bibinfo
  {author} {\bibfnamefont {D.}~\bibnamefont {Weisshaar}}, \bibinfo {author}
  {\bibfnamefont {K.}~\bibnamefont {Whitmore}}, \bibinfo {author}
  {\bibfnamefont {A.}~\bibnamefont {Wiens}}, \bibinfo {author} {\bibfnamefont
  {S.~J.}\ \bibnamefont {Williams}}, \bibinfo {author} {\bibfnamefont
  {K.}~\bibnamefont {Wimmer}}, \ and\ \bibinfo {author} {\bibfnamefont
  {T.}~\bibnamefont {Yamamato}},\ }\href {\doibase 10.1103/PhysRevC.93.031303}
  {\bibfield  {journal} {\bibinfo  {journal} {Phys. Rev. C}\ }\textbf {\bibinfo
  {volume} {93}},\ \bibinfo {pages} {031303} (\bibinfo {year}
  {2016})}\BibitemShut {NoStop}%
\bibitem [{\citenamefont {Terry}\ \emph {et~al.}(2008)\citenamefont {Terry},
  \citenamefont {Brown}, \citenamefont {Campbell}, \citenamefont {Cook},
  \citenamefont {Davies}, \citenamefont {Dinca}, \citenamefont {Gade},
  \citenamefont {Glasmacher}, \citenamefont {Hansen}, \citenamefont {Sherrill},
  \citenamefont {Zwahlen}, \citenamefont {Bazin}, \citenamefont {Yoneda},
  \citenamefont {Tostevin}, \citenamefont {Otsuka}, \citenamefont {Utsuno},\
  and\ \citenamefont {Pritychenko}}]{Terry08}%
  \BibitemOpen
  \bibfield  {author} {\bibinfo {author} {\bibfnamefont {J.~R.}\ \bibnamefont
  {Terry}}, \bibinfo {author} {\bibfnamefont {B.~A.}\ \bibnamefont {Brown}},
  \bibinfo {author} {\bibfnamefont {C.~M.}\ \bibnamefont {Campbell}}, \bibinfo
  {author} {\bibfnamefont {J.~M.}\ \bibnamefont {Cook}}, \bibinfo {author}
  {\bibfnamefont {A.~D.}\ \bibnamefont {Davies}}, \bibinfo {author}
  {\bibfnamefont {D.-C.}\ \bibnamefont {Dinca}}, \bibinfo {author}
  {\bibfnamefont {A.}~\bibnamefont {Gade}}, \bibinfo {author} {\bibfnamefont
  {T.}~\bibnamefont {Glasmacher}}, \bibinfo {author} {\bibfnamefont {P.~G.}\
  \bibnamefont {Hansen}}, \bibinfo {author} {\bibfnamefont {B.~M.}\
  \bibnamefont {Sherrill}}, \bibinfo {author} {\bibfnamefont {H.}~\bibnamefont
  {Zwahlen}}, \bibinfo {author} {\bibfnamefont {D.}~\bibnamefont {Bazin}},
  \bibinfo {author} {\bibfnamefont {K.}~\bibnamefont {Yoneda}}, \bibinfo
  {author} {\bibfnamefont {J.~A.}\ \bibnamefont {Tostevin}}, \bibinfo {author}
  {\bibfnamefont {T.}~\bibnamefont {Otsuka}}, \bibinfo {author} {\bibfnamefont
  {Y.}~\bibnamefont {Utsuno}}, \ and\ \bibinfo {author} {\bibfnamefont
  {B.}~\bibnamefont {Pritychenko}},\ }\href {\doibase
  10.1103/PhysRevC.77.014316} {\bibfield  {journal} {\bibinfo  {journal} {Phys.
  Rev. C}\ }\textbf {\bibinfo {volume} {77}},\ \bibinfo {pages} {014316}
  (\bibinfo {year} {2008})}\BibitemShut {NoStop}%
\bibitem [{\citenamefont {Matoba}\ \emph {et~al.}(1993)\citenamefont {Matoba},
  \citenamefont {Iwamoto}, \citenamefont {Uozumi}, \citenamefont {Sakae},
  \citenamefont {Koori}, \citenamefont {Fujiki}, \citenamefont {Ohgaki},
  \citenamefont {Ijiri}, \citenamefont {Maki},\ and\ \citenamefont
  {Nakano}}]{Mato93}%
  \BibitemOpen
  \bibfield  {author} {\bibinfo {author} {\bibfnamefont {M.}~\bibnamefont
  {Matoba}}, \bibinfo {author} {\bibfnamefont {O.}~\bibnamefont {Iwamoto}},
  \bibinfo {author} {\bibfnamefont {Y.}~\bibnamefont {Uozumi}}, \bibinfo
  {author} {\bibfnamefont {T.}~\bibnamefont {Sakae}}, \bibinfo {author}
  {\bibfnamefont {N.}~\bibnamefont {Koori}}, \bibinfo {author} {\bibfnamefont
  {T.}~\bibnamefont {Fujiki}}, \bibinfo {author} {\bibfnamefont
  {H.}~\bibnamefont {Ohgaki}}, \bibinfo {author} {\bibfnamefont
  {H.}~\bibnamefont {Ijiri}}, \bibinfo {author} {\bibfnamefont
  {T.}~\bibnamefont {Maki}}, \ and\ \bibinfo {author} {\bibfnamefont
  {M.}~\bibnamefont {Nakano}},\ }\href {\doibase 10.1103/PhysRevC.48.95}
  {\bibfield  {journal} {\bibinfo  {journal} {Phys. Rev. C}\ }\textbf {\bibinfo
  {volume} {48}},\ \bibinfo {pages} {95} (\bibinfo {year} {1993})}\BibitemShut
  {NoStop}%
\bibitem [{\citenamefont {Fifield}\ \emph {et~al.}(1985)\citenamefont
  {Fifield}, \citenamefont {Woods}, \citenamefont {Bark}, \citenamefont
  {Drumm},\ and\ \citenamefont {Hotchkis}}]{Fifi85}%
  \BibitemOpen
  \bibfield  {author} {\bibinfo {author} {\bibfnamefont {L.}~\bibnamefont
  {Fifield}}, \bibinfo {author} {\bibfnamefont {C.}~\bibnamefont {Woods}},
  \bibinfo {author} {\bibfnamefont {R.}~\bibnamefont {Bark}}, \bibinfo {author}
  {\bibfnamefont {P.}~\bibnamefont {Drumm}}, \ and\ \bibinfo {author}
  {\bibfnamefont {M.}~\bibnamefont {Hotchkis}},\ }\href {\doibase
  https://doi.org/10.1016/0375-9474(85)90244-1} {\bibfield  {journal} {\bibinfo
   {journal} {Nuclear Physics A}\ }\textbf {\bibinfo {volume} {440}},\ \bibinfo
  {pages} {531 } (\bibinfo {year} {1985})}\BibitemShut {NoStop}%
\bibitem [{\citenamefont {Morton}(2002)}]{Morton}%
  \BibitemOpen
  \bibfield  {author} {\bibinfo {author} {\bibfnamefont {A.~C.}\ \bibnamefont
  {Morton}},\ }\href@noop {} {\bibfield  {journal} {\bibinfo  {journal} {Phys.
  Lett. B}\ }\textbf {\bibinfo {volume} {544}} (\bibinfo {year}
  {2002})}\BibitemShut {NoStop}%
\bibitem [{\citenamefont {Wang}\ \emph {et~al.}(2010)\citenamefont {Wang},
  \citenamefont {Chapman}, \citenamefont {Liang}, \citenamefont {Haas},
  \citenamefont {Bouhelal}, \citenamefont {Azaiez}, \citenamefont {Behera},
  \citenamefont {Burns}, \citenamefont {Caurier}, \citenamefont {Corradi},
  \citenamefont {Curien}, \citenamefont {Deacon}, \citenamefont {Dombr\'adi},
  \citenamefont {Farnea}, \citenamefont {Fioretto}, \citenamefont {Gadea},
  \citenamefont {Hodsdon}, \citenamefont {Ibrahim}, \citenamefont {Jungclaus},
  \citenamefont {Keyes}, \citenamefont {Kumar}, \citenamefont {Latina},
  \citenamefont {M\ifmmode~\u{a}\else \u{a}\fi{}rginean}, \citenamefont
  {Montagnoli}, \citenamefont {Napoli}, \citenamefont {Nowacki}, \citenamefont
  {Ollier}, \citenamefont {O'Donnell}, \citenamefont {Papenberg}, \citenamefont
  {Pollarolo}, \citenamefont {Salsac}, \citenamefont {Scarlassara},
  \citenamefont {Smith}, \citenamefont {Spohr}, \citenamefont {Stanoiu},
  \citenamefont {Stefanini}, \citenamefont {Szilner}, \citenamefont {Trotta},\
  and\ \citenamefont {Verney}}]{Wang10}%
  \BibitemOpen
  \bibfield  {author} {\bibinfo {author} {\bibfnamefont {Z.~M.}\ \bibnamefont
  {Wang}}, \bibinfo {author} {\bibfnamefont {R.}~\bibnamefont {Chapman}},
  \bibinfo {author} {\bibfnamefont {X.}~\bibnamefont {Liang}}, \bibinfo
  {author} {\bibfnamefont {F.}~\bibnamefont {Haas}}, \bibinfo {author}
  {\bibfnamefont {M.}~\bibnamefont {Bouhelal}}, \bibinfo {author}
  {\bibfnamefont {F.}~\bibnamefont {Azaiez}}, \bibinfo {author} {\bibfnamefont
  {B.~R.}\ \bibnamefont {Behera}}, \bibinfo {author} {\bibfnamefont
  {M.}~\bibnamefont {Burns}}, \bibinfo {author} {\bibfnamefont
  {E.}~\bibnamefont {Caurier}}, \bibinfo {author} {\bibfnamefont
  {L.}~\bibnamefont {Corradi}}, \bibinfo {author} {\bibfnamefont
  {D.}~\bibnamefont {Curien}}, \bibinfo {author} {\bibfnamefont {A.~N.}\
  \bibnamefont {Deacon}}, \bibinfo {author} {\bibfnamefont {Z.}~\bibnamefont
  {Dombr\'adi}}, \bibinfo {author} {\bibfnamefont {E.}~\bibnamefont {Farnea}},
  \bibinfo {author} {\bibfnamefont {E.}~\bibnamefont {Fioretto}}, \bibinfo
  {author} {\bibfnamefont {A.}~\bibnamefont {Gadea}}, \bibinfo {author}
  {\bibfnamefont {A.}~\bibnamefont {Hodsdon}}, \bibinfo {author} {\bibfnamefont
  {F.}~\bibnamefont {Ibrahim}}, \bibinfo {author} {\bibfnamefont
  {A.}~\bibnamefont {Jungclaus}}, \bibinfo {author} {\bibfnamefont
  {K.}~\bibnamefont {Keyes}}, \bibinfo {author} {\bibfnamefont
  {V.}~\bibnamefont {Kumar}}, \bibinfo {author} {\bibfnamefont
  {A.}~\bibnamefont {Latina}}, \bibinfo {author} {\bibfnamefont
  {N.}~\bibnamefont {M\ifmmode~\u{a}\else \u{a}\fi{}rginean}}, \bibinfo
  {author} {\bibfnamefont {G.}~\bibnamefont {Montagnoli}}, \bibinfo {author}
  {\bibfnamefont {D.~R.}\ \bibnamefont {Napoli}}, \bibinfo {author}
  {\bibfnamefont {F.}~\bibnamefont {Nowacki}}, \bibinfo {author} {\bibfnamefont
  {J.}~\bibnamefont {Ollier}}, \bibinfo {author} {\bibfnamefont
  {D.}~\bibnamefont {O'Donnell}}, \bibinfo {author} {\bibfnamefont
  {A.}~\bibnamefont {Papenberg}}, \bibinfo {author} {\bibfnamefont
  {G.}~\bibnamefont {Pollarolo}}, \bibinfo {author} {\bibfnamefont {M.-D.}\
  \bibnamefont {Salsac}}, \bibinfo {author} {\bibfnamefont {F.}~\bibnamefont
  {Scarlassara}}, \bibinfo {author} {\bibfnamefont {J.~F.}\ \bibnamefont
  {Smith}}, \bibinfo {author} {\bibfnamefont {K.~M.}\ \bibnamefont {Spohr}},
  \bibinfo {author} {\bibfnamefont {M.}~\bibnamefont {Stanoiu}}, \bibinfo
  {author} {\bibfnamefont {A.~M.}\ \bibnamefont {Stefanini}}, \bibinfo {author}
  {\bibfnamefont {S.}~\bibnamefont {Szilner}}, \bibinfo {author} {\bibfnamefont
  {M.}~\bibnamefont {Trotta}}, \ and\ \bibinfo {author} {\bibfnamefont
  {D.}~\bibnamefont {Verney}},\ }\href {\doibase 10.1103/PhysRevC.81.064301}
  {\bibfield  {journal} {\bibinfo  {journal} {Phys. Rev. C}\ }\textbf {\bibinfo
  {volume} {81}},\ \bibinfo {pages} {064301} (\bibinfo {year}
  {2010})}\BibitemShut {NoStop}%
\bibitem [{\citenamefont {Enders}\ \emph {et~al.}(2002)\citenamefont {Enders},
  \citenamefont {Bauer}, \citenamefont {Bazin}, \citenamefont {Bonaccorso},
  \citenamefont {Brown}, \citenamefont {Glasmacher}, \citenamefont {Hansen},
  \citenamefont {Maddalena}, \citenamefont {Miller}, \citenamefont {Navin},
  \citenamefont {Sherrill},\ and\ \citenamefont {Tostevin}}]{Ende12}%
  \BibitemOpen
  \bibfield  {author} {\bibinfo {author} {\bibfnamefont {J.}~\bibnamefont
  {Enders}}, \bibinfo {author} {\bibfnamefont {A.}~\bibnamefont {Bauer}},
  \bibinfo {author} {\bibfnamefont {D.}~\bibnamefont {Bazin}}, \bibinfo
  {author} {\bibfnamefont {A.}~\bibnamefont {Bonaccorso}}, \bibinfo {author}
  {\bibfnamefont {B.~A.}\ \bibnamefont {Brown}}, \bibinfo {author}
  {\bibfnamefont {T.}~\bibnamefont {Glasmacher}}, \bibinfo {author}
  {\bibfnamefont {P.~G.}\ \bibnamefont {Hansen}}, \bibinfo {author}
  {\bibfnamefont {V.}~\bibnamefont {Maddalena}}, \bibinfo {author}
  {\bibfnamefont {K.~L.}\ \bibnamefont {Miller}}, \bibinfo {author}
  {\bibfnamefont {A.}~\bibnamefont {Navin}}, \bibinfo {author} {\bibfnamefont
  {B.~M.}\ \bibnamefont {Sherrill}}, \ and\ \bibinfo {author} {\bibfnamefont
  {J.~A.}\ \bibnamefont {Tostevin}},\ }\href {\doibase
  10.1103/PhysRevC.65.034318} {\bibfield  {journal} {\bibinfo  {journal} {Phys.
  Rev. C}\ }\textbf {\bibinfo {volume} {65}},\ \bibinfo {pages} {034318}
  (\bibinfo {year} {2002})}\BibitemShut {NoStop}%
\bibitem [{\citenamefont {Weisshaar}\ \emph {et~al.}(2017)\citenamefont
  {Weisshaar}, \citenamefont {Bazin}, \citenamefont {Bender}, \citenamefont
  {Campbell}, \citenamefont {Recchia}, \citenamefont {Bader}, \citenamefont
  {Baugher}, \citenamefont {Belarge}, \citenamefont {Carpenter}, \citenamefont
  {Crawford}, \citenamefont {Cromaz}, \citenamefont {Elman}, \citenamefont
  {Fallon}, \citenamefont {Forney}, \citenamefont {Gade}, \citenamefont
  {Harker}, \citenamefont {Kobayashi}, \citenamefont {Langer}, \citenamefont
  {Lauritsen}, \citenamefont {Lee}, \citenamefont {Lemasson}, \citenamefont
  {Longfellow}, \citenamefont {Lunderberg}, \citenamefont {Macchiavelli},
  \citenamefont {Miki}, \citenamefont {Momiyama}, \citenamefont {Noji},
  \citenamefont {Radford}, \citenamefont {Scott}, \citenamefont {Sethi},
  \citenamefont {Stroberg}, \citenamefont {Sullivan}, \citenamefont {Titus},
  \citenamefont {Wiens}, \citenamefont {Williams}, \citenamefont {Wimmer},\
  and\ \citenamefont {Zhu}}]{Weis17}%
  \BibitemOpen
  \bibfield  {author} {\bibinfo {author} {\bibfnamefont {D.}~\bibnamefont
  {Weisshaar}}, \bibinfo {author} {\bibfnamefont {D.}~\bibnamefont {Bazin}},
  \bibinfo {author} {\bibfnamefont {P.}~\bibnamefont {Bender}}, \bibinfo
  {author} {\bibfnamefont {C.}~\bibnamefont {Campbell}}, \bibinfo {author}
  {\bibfnamefont {F.}~\bibnamefont {Recchia}}, \bibinfo {author} {\bibfnamefont
  {V.}~\bibnamefont {Bader}}, \bibinfo {author} {\bibfnamefont
  {T.}~\bibnamefont {Baugher}}, \bibinfo {author} {\bibfnamefont
  {J.}~\bibnamefont {Belarge}}, \bibinfo {author} {\bibfnamefont
  {M.}~\bibnamefont {Carpenter}}, \bibinfo {author} {\bibfnamefont
  {H.}~\bibnamefont {Crawford}}, \bibinfo {author} {\bibfnamefont
  {M.}~\bibnamefont {Cromaz}}, \bibinfo {author} {\bibfnamefont
  {B.}~\bibnamefont {Elman}}, \bibinfo {author} {\bibfnamefont
  {P.}~\bibnamefont {Fallon}}, \bibinfo {author} {\bibfnamefont
  {A.}~\bibnamefont {Forney}}, \bibinfo {author} {\bibfnamefont
  {A.}~\bibnamefont {Gade}}, \bibinfo {author} {\bibfnamefont {J.}~\bibnamefont
  {Harker}}, \bibinfo {author} {\bibfnamefont {N.}~\bibnamefont {Kobayashi}},
  \bibinfo {author} {\bibfnamefont {C.}~\bibnamefont {Langer}}, \bibinfo
  {author} {\bibfnamefont {T.}~\bibnamefont {Lauritsen}}, \bibinfo {author}
  {\bibfnamefont {I.}~\bibnamefont {Lee}}, \bibinfo {author} {\bibfnamefont
  {A.}~\bibnamefont {Lemasson}}, \bibinfo {author} {\bibfnamefont
  {B.}~\bibnamefont {Longfellow}}, \bibinfo {author} {\bibfnamefont
  {E.}~\bibnamefont {Lunderberg}}, \bibinfo {author} {\bibfnamefont
  {A.}~\bibnamefont {Macchiavelli}}, \bibinfo {author} {\bibfnamefont
  {K.}~\bibnamefont {Miki}}, \bibinfo {author} {\bibfnamefont {S.}~\bibnamefont
  {Momiyama}}, \bibinfo {author} {\bibfnamefont {S.}~\bibnamefont {Noji}},
  \bibinfo {author} {\bibfnamefont {D.}~\bibnamefont {Radford}}, \bibinfo
  {author} {\bibfnamefont {M.}~\bibnamefont {Scott}}, \bibinfo {author}
  {\bibfnamefont {J.}~\bibnamefont {Sethi}}, \bibinfo {author} {\bibfnamefont
  {S.}~\bibnamefont {Stroberg}}, \bibinfo {author} {\bibfnamefont
  {C.}~\bibnamefont {Sullivan}}, \bibinfo {author} {\bibfnamefont
  {R.}~\bibnamefont {Titus}}, \bibinfo {author} {\bibfnamefont
  {A.}~\bibnamefont {Wiens}}, \bibinfo {author} {\bibfnamefont
  {S.}~\bibnamefont {Williams}}, \bibinfo {author} {\bibfnamefont
  {K.}~\bibnamefont {Wimmer}}, \ and\ \bibinfo {author} {\bibfnamefont
  {S.}~\bibnamefont {Zhu}},\ }\href {\doibase
  https://doi.org/10.1016/j.nima.2016.12.001} {\bibfield  {journal} {\bibinfo
  {journal} {Nucl. Instrum. Methods Phys. Res. A}\ }\textbf {\bibinfo {volume}
  {847}},\ \bibinfo {pages} {187 } (\bibinfo {year} {2017})}\BibitemShut
  {NoStop}%
\bibitem [{\citenamefont {Paschalis}\ \emph {et~al.}(2013)\citenamefont
  {Paschalis}, \citenamefont {Lee}, \citenamefont {Macchiavelli}, \citenamefont
  {Campbell}, \citenamefont {Cromaz}, \citenamefont {Gros}, \citenamefont
  {Pavan}, \citenamefont {Qian}, \citenamefont {Clark}, \citenamefont
  {Crawford}, \citenamefont {Doering}, \citenamefont {Fallon}, \citenamefont
  {Lionberger}, \citenamefont {Loew}, \citenamefont {Petri}, \citenamefont
  {Stezelberger}, \citenamefont {Zimmermann}, \citenamefont {Radford},
  \citenamefont {Lagergren}, \citenamefont {Winkler}, \citenamefont
  {Glasmacher}, \citenamefont {Anderson},\ and\ \citenamefont
  {Beausang}}]{Pasc13}%
  \BibitemOpen
  \bibfield  {author} {\bibinfo {author} {\bibfnamefont {S.}~\bibnamefont
  {Paschalis}}, \bibinfo {author} {\bibfnamefont {I.~Y.}\ \bibnamefont {Lee}},
  \bibinfo {author} {\bibfnamefont {A.~O.}\ \bibnamefont {Macchiavelli}},
  \bibinfo {author} {\bibfnamefont {C.~M.}\ \bibnamefont {Campbell}}, \bibinfo
  {author} {\bibfnamefont {M.}~\bibnamefont {Cromaz}}, \bibinfo {author}
  {\bibfnamefont {S.}~\bibnamefont {Gros}}, \bibinfo {author} {\bibfnamefont
  {J.}~\bibnamefont {Pavan}}, \bibinfo {author} {\bibfnamefont
  {J.}~\bibnamefont {Qian}}, \bibinfo {author} {\bibfnamefont {R.~M.}\
  \bibnamefont {Clark}}, \bibinfo {author} {\bibfnamefont {H.~L.}\ \bibnamefont
  {Crawford}}, \bibinfo {author} {\bibfnamefont {D.}~\bibnamefont {Doering}},
  \bibinfo {author} {\bibfnamefont {P.}~\bibnamefont {Fallon}}, \bibinfo
  {author} {\bibfnamefont {C.}~\bibnamefont {Lionberger}}, \bibinfo {author}
  {\bibfnamefont {T.}~\bibnamefont {Loew}}, \bibinfo {author} {\bibfnamefont
  {M.}~\bibnamefont {Petri}}, \bibinfo {author} {\bibfnamefont
  {T.}~\bibnamefont {Stezelberger}}, \bibinfo {author} {\bibfnamefont
  {S.}~\bibnamefont {Zimmermann}}, \bibinfo {author} {\bibfnamefont {D.~C.}\
  \bibnamefont {Radford}}, \bibinfo {author} {\bibfnamefont {D.}~\bibnamefont
  {Lagergren}, \bibfnamefont {K.~Weisshaar}}, \bibinfo {author} {\bibfnamefont
  {R.}~\bibnamefont {Winkler}}, \bibinfo {author} {\bibfnamefont
  {T.}~\bibnamefont {Glasmacher}}, \bibinfo {author} {\bibfnamefont {J.~T.}\
  \bibnamefont {Anderson}}, \ and\ \bibinfo {author} {\bibfnamefont {C.~W.}\
  \bibnamefont {Beausang}},\ }\href
  {https://dx.doi.org/10.1016/j.nima.2013.01.009} {\bibfield  {journal}
  {\bibinfo  {journal} {Nucl. Instrum. Methods Phys. Res. A}\ }\textbf
  {\bibinfo {volume} {709}},\ \bibinfo {pages} {44} (\bibinfo {year}
  {2013})}\BibitemShut {NoStop}%
\bibitem [{\citenamefont {Morrissey}\ \emph {et~al.}(2003)\citenamefont
  {Morrissey}, \citenamefont {Sherrill}, \citenamefont {Steiner}, \citenamefont
  {Stolz},\ and\ \citenamefont {Wiedenhoever}}]{Morrisey}%
  \BibitemOpen
  \bibfield  {author} {\bibinfo {author} {\bibfnamefont {D.~J.}\ \bibnamefont
  {Morrissey}}, \bibinfo {author} {\bibfnamefont {B.~M.}\ \bibnamefont
  {Sherrill}}, \bibinfo {author} {\bibfnamefont {M.}~\bibnamefont {Steiner}},
  \bibinfo {author} {\bibfnamefont {A.}~\bibnamefont {Stolz}}, \ and\ \bibinfo
  {author} {\bibfnamefont {I.}~\bibnamefont {Wiedenhoever}},\ }\href
  {https://dx.doi.org/10.1016/S0168-583X(02)01895-5} {\bibfield  {journal}
  {\bibinfo  {journal} {Nucl.Instrum. Methods Phys. Res. B}\ }\textbf {\bibinfo
  {volume} {204}},\ \bibinfo {pages} {90} (\bibinfo {year} {2003})}\BibitemShut
  {NoStop}%
\bibitem [{\citenamefont {Bazin}\ \emph {et~al.}(2003)\citenamefont {Bazin},
  \citenamefont {Caggiano}, \citenamefont {Sherrill}, \citenamefont {Yurkon},\
  and\ \citenamefont {Zeller}}]{Bazin2}%
  \BibitemOpen
  \bibfield  {author} {\bibinfo {author} {\bibfnamefont {D.}~\bibnamefont
  {Bazin}}, \bibinfo {author} {\bibfnamefont {J.}~\bibnamefont {Caggiano}},
  \bibinfo {author} {\bibfnamefont {B.}~\bibnamefont {Sherrill}}, \bibinfo
  {author} {\bibfnamefont {J.}~\bibnamefont {Yurkon}}, \ and\ \bibinfo {author}
  {\bibfnamefont {A.}~\bibnamefont {Zeller}},\ }\href
  {https://dx.doi.org/10.1016/S0168-583X(02)02142-0} {\bibfield  {journal}
  {\bibinfo  {journal} {Nucl. Instrum. Methods Phys. Res. A}\ }\textbf
  {\bibinfo {volume} {204}},\ \bibinfo {pages} {629} (\bibinfo {year}
  {2003})}\BibitemShut {NoStop}%
\bibitem [{\citenamefont {Bazin}\ \emph {et~al.}(1999)\citenamefont {Bazin},
  \citenamefont {Benenson}, \citenamefont {Morrissey}, \citenamefont
  {Sherrill}, \citenamefont {Swan},\ and\ \citenamefont {Swanson}}]{Bazin}%
  \BibitemOpen
  \bibfield  {author} {\bibinfo {author} {\bibfnamefont {D.}~\bibnamefont
  {Bazin}}, \bibinfo {author} {\bibfnamefont {W.}~\bibnamefont {Benenson}},
  \bibinfo {author} {\bibfnamefont {D.~J.}\ \bibnamefont {Morrissey}}, \bibinfo
  {author} {\bibfnamefont {B.}~\bibnamefont {Sherrill}}, \bibinfo {author}
  {\bibfnamefont {D.}~\bibnamefont {Swan}}, \ and\ \bibinfo {author}
  {\bibfnamefont {R.}~\bibnamefont {Swanson}},\ }\href
  {https://dx.doi.org/10.1016/S0168-9002(98)00960-7} {\bibfield  {journal}
  {\bibinfo  {journal} {Nucl. Instrum. Methods Phys. Res. A}\ }\textbf
  {\bibinfo {volume} {422}},\ \bibinfo {pages} {291} (\bibinfo {year}
  {1999})}\BibitemShut {NoStop}%
\bibitem [{\citenamefont {Berz}\ \emph {et~al.}(1993)\citenamefont {Berz},
  \citenamefont {Joh}, \citenamefont {Nolen}, \citenamefont {Sherrill},\ and\
  \citenamefont {Zeller}}]{COSY}%
  \BibitemOpen
  \bibfield  {author} {\bibinfo {author} {\bibfnamefont {M.}~\bibnamefont
  {Berz}}, \bibinfo {author} {\bibfnamefont {K.}~\bibnamefont {Joh}}, \bibinfo
  {author} {\bibfnamefont {J.~A.}\ \bibnamefont {Nolen}}, \bibinfo {author}
  {\bibfnamefont {B.~M.}\ \bibnamefont {Sherrill}}, \ and\ \bibinfo {author}
  {\bibfnamefont {A.~F.}\ \bibnamefont {Zeller}},\ }\href
  {https://dx.doi.org/10.1103/PhysRevC.47.537} {\bibfield  {journal} {\bibinfo
  {journal} {Phys. Rev. C}\ }\textbf {\bibinfo {volume} {47}},\ \bibinfo
  {pages} {537} (\bibinfo {year} {1993})}\BibitemShut {NoStop}%
\bibitem [{\citenamefont {Agostinelli}\ \emph {et~al.}(2003)\citenamefont
  {Agostinelli}, \citenamefont {Allison}, \citenamefont {Amako}, \citenamefont
  {Apostolakis}, \citenamefont {Araujo}, \citenamefont {Arce}, \citenamefont
  {Asai}, \citenamefont {Axen}, \citenamefont {Banerjee}, \citenamefont
  {Barrand}, \citenamefont {Behner}, \citenamefont {Bellagamba}, \citenamefont
  {Boudreau}, \citenamefont {Broglia}, \citenamefont {Brunengo}, \citenamefont
  {Burkhardt}, \citenamefont {Chauvie}, \citenamefont {Chuma}, \citenamefont
  {Chytracek}, \citenamefont {Cooperman}, \citenamefont {Cosmo}, \citenamefont
  {Degtyarenko}, \citenamefont {Dell'Acqua}, \citenamefont {Depaola},
  \citenamefont {Dietrich}, \citenamefont {Enami}, \citenamefont {Feliciello},
  \citenamefont {Ferguson}, \citenamefont {Fesefeldt}, \citenamefont {Folger},
  \citenamefont {Foppiano}, \citenamefont {Forti}, \citenamefont {Garelli},
  \citenamefont {Giani}, \citenamefont {Giannitrapani}, \citenamefont {Gibin},
  \citenamefont {Cadenas}, \citenamefont {González}, \citenamefont {Abril},
  \citenamefont {Greeniaus}, \citenamefont {Greiner}, \citenamefont {Grichine},
  \citenamefont {Grossheim}, \citenamefont {Guatelli}, \citenamefont
  {Gumplinger}, \citenamefont {Hamatsu}, \citenamefont {Hashimoto},
  \citenamefont {Hasui}, \citenamefont {Heikkinen}, \citenamefont {Howard},
  \citenamefont {Ivanchenko}, \citenamefont {Johnson}, \citenamefont {Jones},
  \citenamefont {Kallenbach}, \citenamefont {Kanaya}, \citenamefont {Kawabata},
  \citenamefont {Kawabata}, \citenamefont {Kawaguti}, \citenamefont {Kelner},
  \citenamefont {Kent}, \citenamefont {Kimura}, \citenamefont {Kodama},
  \citenamefont {Kokoulin}, \citenamefont {Kossov}, \citenamefont {Kurashige},
  \citenamefont {Lamanna}, \citenamefont {Lampén}, \citenamefont {Lara},
  \citenamefont {Lefebure}, \citenamefont {Lei}, \citenamefont {Liendl},
  \citenamefont {Lockman}, \citenamefont {Longo}, \citenamefont {Magni},
  \citenamefont {Maire}, \citenamefont {Medernach}, \citenamefont {Minamimoto},
  \citenamefont {de~Freitas}, \citenamefont {Morita}, \citenamefont {Murakami},
  \citenamefont {Nagamatu}, \citenamefont {Nartallo}, \citenamefont {Nieminen},
  \citenamefont {Nishimura}, \citenamefont {Ohtsubo}, \citenamefont {Okamura},
  \citenamefont {O'Neale}, \citenamefont {Oohata}, \citenamefont {Paech},
  \citenamefont {Perl}, \citenamefont {Pfeiffer}, \citenamefont {Pia},
  \citenamefont {Ranjard}, \citenamefont {Rybin}, \citenamefont {Sadilov},
  \citenamefont {Salvo}, \citenamefont {Santin}, \citenamefont {Sasaki},
  \citenamefont {Savvas}, \citenamefont {Sawada}, \citenamefont {Scherer},
  \citenamefont {Sei}, \citenamefont {Sirotenko}, \citenamefont {Smith},
  \citenamefont {Starkov}, \citenamefont {Stoecker}, \citenamefont {Sulkimo},
  \citenamefont {Takahata}, \citenamefont {Tanaka}, \citenamefont {Tcherniaev},
  \citenamefont {Tehrani}, \citenamefont {Tropeano}, \citenamefont {Truscott},
  \citenamefont {Uno}, \citenamefont {Urban}, \citenamefont {Urban},
  \citenamefont {Verderi}, \citenamefont {Walkden}, \citenamefont {Wander},
  \citenamefont {Weber}, \citenamefont {Wellisch}, \citenamefont {Wenaus},
  \citenamefont {Williams}, \citenamefont {Wright}, \citenamefont {Yamada},
  \citenamefont {Yoshida},\ and\ \citenamefont {Zschiesche}}]{Agos03}%
  \BibitemOpen
  \bibfield  {author} {\bibinfo {author} {\bibfnamefont {S.}~\bibnamefont
  {Agostinelli}}, \bibinfo {author} {\bibfnamefont {J.}~\bibnamefont
  {Allison}}, \bibinfo {author} {\bibfnamefont {K.}~\bibnamefont {Amako}},
  \bibinfo {author} {\bibfnamefont {J.}~\bibnamefont {Apostolakis}}, \bibinfo
  {author} {\bibfnamefont {H.}~\bibnamefont {Araujo}}, \bibinfo {author}
  {\bibfnamefont {P.}~\bibnamefont {Arce}}, \bibinfo {author} {\bibfnamefont
  {M.}~\bibnamefont {Asai}}, \bibinfo {author} {\bibfnamefont {D.}~\bibnamefont
  {Axen}}, \bibinfo {author} {\bibfnamefont {S.}~\bibnamefont {Banerjee}},
  \bibinfo {author} {\bibfnamefont {G.}~\bibnamefont {Barrand}}, \bibinfo
  {author} {\bibfnamefont {F.}~\bibnamefont {Behner}}, \bibinfo {author}
  {\bibfnamefont {L.}~\bibnamefont {Bellagamba}}, \bibinfo {author}
  {\bibfnamefont {J.}~\bibnamefont {Boudreau}}, \bibinfo {author}
  {\bibfnamefont {L.}~\bibnamefont {Broglia}}, \bibinfo {author} {\bibfnamefont
  {A.}~\bibnamefont {Brunengo}}, \bibinfo {author} {\bibfnamefont
  {H.}~\bibnamefont {Burkhardt}}, \bibinfo {author} {\bibfnamefont
  {S.}~\bibnamefont {Chauvie}}, \bibinfo {author} {\bibfnamefont
  {J.}~\bibnamefont {Chuma}}, \bibinfo {author} {\bibfnamefont
  {R.}~\bibnamefont {Chytracek}}, \bibinfo {author} {\bibfnamefont
  {G.}~\bibnamefont {Cooperman}}, \bibinfo {author} {\bibfnamefont
  {G.}~\bibnamefont {Cosmo}}, \bibinfo {author} {\bibfnamefont
  {P.}~\bibnamefont {Degtyarenko}}, \bibinfo {author} {\bibfnamefont
  {A.}~\bibnamefont {Dell'Acqua}}, \bibinfo {author} {\bibfnamefont
  {G.}~\bibnamefont {Depaola}}, \bibinfo {author} {\bibfnamefont
  {D.}~\bibnamefont {Dietrich}}, \bibinfo {author} {\bibfnamefont
  {R.}~\bibnamefont {Enami}}, \bibinfo {author} {\bibfnamefont
  {A.}~\bibnamefont {Feliciello}}, \bibinfo {author} {\bibfnamefont
  {C.}~\bibnamefont {Ferguson}}, \bibinfo {author} {\bibfnamefont
  {H.}~\bibnamefont {Fesefeldt}}, \bibinfo {author} {\bibfnamefont
  {G.}~\bibnamefont {Folger}}, \bibinfo {author} {\bibfnamefont
  {F.}~\bibnamefont {Foppiano}}, \bibinfo {author} {\bibfnamefont
  {A.}~\bibnamefont {Forti}}, \bibinfo {author} {\bibfnamefont
  {S.}~\bibnamefont {Garelli}}, \bibinfo {author} {\bibfnamefont
  {S.}~\bibnamefont {Giani}}, \bibinfo {author} {\bibfnamefont
  {R.}~\bibnamefont {Giannitrapani}}, \bibinfo {author} {\bibfnamefont
  {D.}~\bibnamefont {Gibin}}, \bibinfo {author} {\bibfnamefont {J.~G.}\
  \bibnamefont {Cadenas}}, \bibinfo {author} {\bibfnamefont {I.}~\bibnamefont
  {González}}, \bibinfo {author} {\bibfnamefont {G.~G.}\ \bibnamefont
  {Abril}}, \bibinfo {author} {\bibfnamefont {G.}~\bibnamefont {Greeniaus}},
  \bibinfo {author} {\bibfnamefont {W.}~\bibnamefont {Greiner}}, \bibinfo
  {author} {\bibfnamefont {V.}~\bibnamefont {Grichine}}, \bibinfo {author}
  {\bibfnamefont {A.}~\bibnamefont {Grossheim}}, \bibinfo {author}
  {\bibfnamefont {S.}~\bibnamefont {Guatelli}}, \bibinfo {author}
  {\bibfnamefont {P.}~\bibnamefont {Gumplinger}}, \bibinfo {author}
  {\bibfnamefont {R.}~\bibnamefont {Hamatsu}}, \bibinfo {author} {\bibfnamefont
  {K.}~\bibnamefont {Hashimoto}}, \bibinfo {author} {\bibfnamefont
  {H.}~\bibnamefont {Hasui}}, \bibinfo {author} {\bibfnamefont
  {A.}~\bibnamefont {Heikkinen}}, \bibinfo {author} {\bibfnamefont
  {A.}~\bibnamefont {Howard}}, \bibinfo {author} {\bibfnamefont
  {V.}~\bibnamefont {Ivanchenko}}, \bibinfo {author} {\bibfnamefont
  {A.}~\bibnamefont {Johnson}}, \bibinfo {author} {\bibfnamefont
  {F.}~\bibnamefont {Jones}}, \bibinfo {author} {\bibfnamefont
  {J.}~\bibnamefont {Kallenbach}}, \bibinfo {author} {\bibfnamefont
  {N.}~\bibnamefont {Kanaya}}, \bibinfo {author} {\bibfnamefont
  {M.}~\bibnamefont {Kawabata}}, \bibinfo {author} {\bibfnamefont
  {Y.}~\bibnamefont {Kawabata}}, \bibinfo {author} {\bibfnamefont
  {M.}~\bibnamefont {Kawaguti}}, \bibinfo {author} {\bibfnamefont
  {S.}~\bibnamefont {Kelner}}, \bibinfo {author} {\bibfnamefont
  {P.}~\bibnamefont {Kent}}, \bibinfo {author} {\bibfnamefont {A.}~\bibnamefont
  {Kimura}}, \bibinfo {author} {\bibfnamefont {T.}~\bibnamefont {Kodama}},
  \bibinfo {author} {\bibfnamefont {R.}~\bibnamefont {Kokoulin}}, \bibinfo
  {author} {\bibfnamefont {M.}~\bibnamefont {Kossov}}, \bibinfo {author}
  {\bibfnamefont {H.}~\bibnamefont {Kurashige}}, \bibinfo {author}
  {\bibfnamefont {E.}~\bibnamefont {Lamanna}}, \bibinfo {author} {\bibfnamefont
  {T.}~\bibnamefont {Lampén}}, \bibinfo {author} {\bibfnamefont
  {V.}~\bibnamefont {Lara}}, \bibinfo {author} {\bibfnamefont {V.}~\bibnamefont
  {Lefebure}}, \bibinfo {author} {\bibfnamefont {F.}~\bibnamefont {Lei}},
  \bibinfo {author} {\bibfnamefont {M.}~\bibnamefont {Liendl}}, \bibinfo
  {author} {\bibfnamefont {W.}~\bibnamefont {Lockman}}, \bibinfo {author}
  {\bibfnamefont {F.}~\bibnamefont {Longo}}, \bibinfo {author} {\bibfnamefont
  {S.}~\bibnamefont {Magni}}, \bibinfo {author} {\bibfnamefont
  {M.}~\bibnamefont {Maire}}, \bibinfo {author} {\bibfnamefont
  {E.}~\bibnamefont {Medernach}}, \bibinfo {author} {\bibfnamefont
  {K.}~\bibnamefont {Minamimoto}}, \bibinfo {author} {\bibfnamefont {P.~M.}\
  \bibnamefont {de~Freitas}}, \bibinfo {author} {\bibfnamefont
  {Y.}~\bibnamefont {Morita}}, \bibinfo {author} {\bibfnamefont
  {K.}~\bibnamefont {Murakami}}, \bibinfo {author} {\bibfnamefont
  {M.}~\bibnamefont {Nagamatu}}, \bibinfo {author} {\bibfnamefont
  {R.}~\bibnamefont {Nartallo}}, \bibinfo {author} {\bibfnamefont
  {P.}~\bibnamefont {Nieminen}}, \bibinfo {author} {\bibfnamefont
  {T.}~\bibnamefont {Nishimura}}, \bibinfo {author} {\bibfnamefont
  {K.}~\bibnamefont {Ohtsubo}}, \bibinfo {author} {\bibfnamefont
  {M.}~\bibnamefont {Okamura}}, \bibinfo {author} {\bibfnamefont
  {S.}~\bibnamefont {O'Neale}}, \bibinfo {author} {\bibfnamefont
  {Y.}~\bibnamefont {Oohata}}, \bibinfo {author} {\bibfnamefont
  {K.}~\bibnamefont {Paech}}, \bibinfo {author} {\bibfnamefont
  {J.}~\bibnamefont {Perl}}, \bibinfo {author} {\bibfnamefont {A.}~\bibnamefont
  {Pfeiffer}}, \bibinfo {author} {\bibfnamefont {M.}~\bibnamefont {Pia}},
  \bibinfo {author} {\bibfnamefont {F.}~\bibnamefont {Ranjard}}, \bibinfo
  {author} {\bibfnamefont {A.}~\bibnamefont {Rybin}}, \bibinfo {author}
  {\bibfnamefont {S.}~\bibnamefont {Sadilov}}, \bibinfo {author} {\bibfnamefont
  {E.~D.}\ \bibnamefont {Salvo}}, \bibinfo {author} {\bibfnamefont
  {G.}~\bibnamefont {Santin}}, \bibinfo {author} {\bibfnamefont
  {T.}~\bibnamefont {Sasaki}}, \bibinfo {author} {\bibfnamefont
  {N.}~\bibnamefont {Savvas}}, \bibinfo {author} {\bibfnamefont
  {Y.}~\bibnamefont {Sawada}}, \bibinfo {author} {\bibfnamefont
  {S.}~\bibnamefont {Scherer}}, \bibinfo {author} {\bibfnamefont
  {S.}~\bibnamefont {Sei}}, \bibinfo {author} {\bibfnamefont {V.}~\bibnamefont
  {Sirotenko}}, \bibinfo {author} {\bibfnamefont {D.}~\bibnamefont {Smith}},
  \bibinfo {author} {\bibfnamefont {N.}~\bibnamefont {Starkov}}, \bibinfo
  {author} {\bibfnamefont {H.}~\bibnamefont {Stoecker}}, \bibinfo {author}
  {\bibfnamefont {J.}~\bibnamefont {Sulkimo}}, \bibinfo {author} {\bibfnamefont
  {M.}~\bibnamefont {Takahata}}, \bibinfo {author} {\bibfnamefont
  {S.}~\bibnamefont {Tanaka}}, \bibinfo {author} {\bibfnamefont
  {E.}~\bibnamefont {Tcherniaev}}, \bibinfo {author} {\bibfnamefont {E.~S.}\
  \bibnamefont {Tehrani}}, \bibinfo {author} {\bibfnamefont {M.}~\bibnamefont
  {Tropeano}}, \bibinfo {author} {\bibfnamefont {P.}~\bibnamefont {Truscott}},
  \bibinfo {author} {\bibfnamefont {H.}~\bibnamefont {Uno}}, \bibinfo {author}
  {\bibfnamefont {L.}~\bibnamefont {Urban}}, \bibinfo {author} {\bibfnamefont
  {P.}~\bibnamefont {Urban}}, \bibinfo {author} {\bibfnamefont
  {M.}~\bibnamefont {Verderi}}, \bibinfo {author} {\bibfnamefont
  {A.}~\bibnamefont {Walkden}}, \bibinfo {author} {\bibfnamefont
  {W.}~\bibnamefont {Wander}}, \bibinfo {author} {\bibfnamefont
  {H.}~\bibnamefont {Weber}}, \bibinfo {author} {\bibfnamefont
  {J.}~\bibnamefont {Wellisch}}, \bibinfo {author} {\bibfnamefont
  {T.}~\bibnamefont {Wenaus}}, \bibinfo {author} {\bibfnamefont
  {D.}~\bibnamefont {Williams}}, \bibinfo {author} {\bibfnamefont
  {D.}~\bibnamefont {Wright}}, \bibinfo {author} {\bibfnamefont
  {T.}~\bibnamefont {Yamada}}, \bibinfo {author} {\bibfnamefont
  {H.}~\bibnamefont {Yoshida}}, \ and\ \bibinfo {author} {\bibfnamefont
  {D.}~\bibnamefont {Zschiesche}},\ }\href {\doibase
  https://doi.org/10.1016/S0168-9002(03)01368-8} {\bibfield  {journal}
  {\bibinfo  {journal} {Nucl. Instrum. and Methods Phys. Res. A}\ }\textbf
  {\bibinfo {volume} {506}},\ \bibinfo {pages} {250 } (\bibinfo {year}
  {2003})}\BibitemShut {NoStop}%
\bibitem [{\citenamefont {Mutschler}\ \emph {et~al.}(2016)\citenamefont
  {Mutschler}, \citenamefont {Sorlin}, \citenamefont {Lemasson}, \citenamefont
  {Bazin}, \citenamefont {Borcea}, \citenamefont {Borcea}, \citenamefont
  {Gade}, \citenamefont {Iwasaki}, \citenamefont {Khan}, \citenamefont
  {Lepailleur}, \citenamefont {Recchia}, \citenamefont {Roger}, \citenamefont
  {Rotaru}, \citenamefont {Stanoiu}, \citenamefont {Stroberg}, \citenamefont
  {Tostevin}, \citenamefont {Vandebrouck}, \citenamefont {Weisshaar},\ and\
  \citenamefont {Wimmer}}]{Muts16b}%
  \BibitemOpen
  \bibfield  {author} {\bibinfo {author} {\bibfnamefont {A.}~\bibnamefont
  {Mutschler}}, \bibinfo {author} {\bibfnamefont {O.}~\bibnamefont {Sorlin}},
  \bibinfo {author} {\bibfnamefont {A.}~\bibnamefont {Lemasson}}, \bibinfo
  {author} {\bibfnamefont {D.}~\bibnamefont {Bazin}}, \bibinfo {author}
  {\bibfnamefont {C.}~\bibnamefont {Borcea}}, \bibinfo {author} {\bibfnamefont
  {R.}~\bibnamefont {Borcea}}, \bibinfo {author} {\bibfnamefont
  {A.}~\bibnamefont {Gade}}, \bibinfo {author} {\bibfnamefont {H.}~\bibnamefont
  {Iwasaki}}, \bibinfo {author} {\bibfnamefont {E.}~\bibnamefont {Khan}},
  \bibinfo {author} {\bibfnamefont {A.}~\bibnamefont {Lepailleur}}, \bibinfo
  {author} {\bibfnamefont {F.}~\bibnamefont {Recchia}}, \bibinfo {author}
  {\bibfnamefont {T.}~\bibnamefont {Roger}}, \bibinfo {author} {\bibfnamefont
  {F.}~\bibnamefont {Rotaru}}, \bibinfo {author} {\bibfnamefont
  {M.}~\bibnamefont {Stanoiu}}, \bibinfo {author} {\bibfnamefont {S.~R.}\
  \bibnamefont {Stroberg}}, \bibinfo {author} {\bibfnamefont {J.~A.}\
  \bibnamefont {Tostevin}}, \bibinfo {author} {\bibfnamefont {M.}~\bibnamefont
  {Vandebrouck}}, \bibinfo {author} {\bibfnamefont {D.}~\bibnamefont
  {Weisshaar}}, \ and\ \bibinfo {author} {\bibfnamefont {K.}~\bibnamefont
  {Wimmer}},\ }\href {\doibase 10.1103/PhysRevC.93.034333} {\bibfield
  {journal} {\bibinfo  {journal} {Phys. Rev. C}\ }\textbf {\bibinfo {volume}
  {93}},\ \bibinfo {pages} {034333} (\bibinfo {year} {2016})}\BibitemShut
  {NoStop}%
\bibitem [{Ens()}]{Ensdf}%
  \BibitemOpen
  \href@noop {} {\enquote {\bibinfo {title} {Evaluated nuclear structure data
  file (ensdf)},}\ }\bibinfo {howpublished}
  {\url{https://www.nndc.bnl.gov/ensdf}}\BibitemShut {NoStop}%
\bibitem [{\citenamefont {Hansen}\ and\ \citenamefont
  {Tostevin}(2003)}]{hansen03}%
  \BibitemOpen
  \bibfield  {author} {\bibinfo {author} {\bibfnamefont {P.}~\bibnamefont
  {Hansen}}\ and\ \bibinfo {author} {\bibfnamefont {J.}~\bibnamefont
  {Tostevin}},\ }\href {https://doi.org/10.1146/annurev.nucl.53.041002.110406}
  {\bibfield  {journal} {\bibinfo  {journal} {Annual Review of Nuclear and
  Particle Science}\ }\textbf {\bibinfo {volume} {53}},\ \bibinfo {pages} {219}
  (\bibinfo {year} {2003})}\BibitemShut {NoStop}%
\bibitem [{\citenamefont {Tostevin}\ and\ \citenamefont
  {Gade}(2014)}]{Toste14}%
  \BibitemOpen
  \bibfield  {author} {\bibinfo {author} {\bibfnamefont {J.~A.}\ \bibnamefont
  {Tostevin}}\ and\ \bibinfo {author} {\bibfnamefont {A.}~\bibnamefont
  {Gade}},\ }\href {\doibase 10.1103/PhysRevC.90.057602} {\bibfield  {journal}
  {\bibinfo  {journal} {Phys. Rev. C}\ }\textbf {\bibinfo {volume} {90}},\
  \bibinfo {pages} {057602} (\bibinfo {year} {2014})}\BibitemShut {NoStop}%
\bibitem [{\citenamefont {Gade}\ \emph {et~al.}(2008)\citenamefont {Gade},
  \citenamefont {Adrich}, \citenamefont {Bazin}, \citenamefont {Bowen},
  \citenamefont {Brown}, \citenamefont {Campbell}, \citenamefont {Cook},
  \citenamefont {Glasmacher}, \citenamefont {Hansen}, \citenamefont {Hosier},
  \citenamefont {McDaniel}, \citenamefont {McGlinchery}, \citenamefont
  {Obertelli}, \citenamefont {Siwek}, \citenamefont {Riley}, \citenamefont
  {Tostevin},\ and\ \citenamefont {Weisshaar}}]{Gade08}%
  \BibitemOpen
  \bibfield  {author} {\bibinfo {author} {\bibfnamefont {A.}~\bibnamefont
  {Gade}}, \bibinfo {author} {\bibfnamefont {P.}~\bibnamefont {Adrich}},
  \bibinfo {author} {\bibfnamefont {D.}~\bibnamefont {Bazin}}, \bibinfo
  {author} {\bibfnamefont {M.~D.}\ \bibnamefont {Bowen}}, \bibinfo {author}
  {\bibfnamefont {B.~A.}\ \bibnamefont {Brown}}, \bibinfo {author}
  {\bibfnamefont {C.~M.}\ \bibnamefont {Campbell}}, \bibinfo {author}
  {\bibfnamefont {J.~M.}\ \bibnamefont {Cook}}, \bibinfo {author}
  {\bibfnamefont {T.}~\bibnamefont {Glasmacher}}, \bibinfo {author}
  {\bibfnamefont {P.~G.}\ \bibnamefont {Hansen}}, \bibinfo {author}
  {\bibfnamefont {K.}~\bibnamefont {Hosier}}, \bibinfo {author} {\bibfnamefont
  {S.}~\bibnamefont {McDaniel}}, \bibinfo {author} {\bibfnamefont
  {D.}~\bibnamefont {McGlinchery}}, \bibinfo {author} {\bibfnamefont
  {A.}~\bibnamefont {Obertelli}}, \bibinfo {author} {\bibfnamefont
  {K.}~\bibnamefont {Siwek}}, \bibinfo {author} {\bibfnamefont {L.~A.}\
  \bibnamefont {Riley}}, \bibinfo {author} {\bibfnamefont {J.~A.}\ \bibnamefont
  {Tostevin}}, \ and\ \bibinfo {author} {\bibfnamefont {D.}~\bibnamefont
  {Weisshaar}},\ }\href {\doibase 10.1103/PhysRevC.77.044306} {\bibfield
  {journal} {\bibinfo  {journal} {Phys. Rev. C}\ }\textbf {\bibinfo {volume}
  {77}},\ \bibinfo {pages} {044306} (\bibinfo {year} {2008})}\BibitemShut
  {NoStop}%
\bibitem [{\citenamefont {Brown}(1998)}]{Skyrme}%
  \BibitemOpen
  \bibfield  {author} {\bibinfo {author} {\bibfnamefont {B.~A.}\ \bibnamefont
  {Brown}},\ }\href@noop {} {\bibfield  {journal} {\bibinfo  {journal} {Phys.
  Rev. C}\ }\textbf {\bibinfo {volume} {58}},\ \bibinfo {pages} {220} (\bibinfo
  {year} {1998})}\BibitemShut {NoStop}%
\bibitem [{\citenamefont {Bertulani}\ and\ \citenamefont
  {Hansen}(2004)}]{bertulani04}%
  \BibitemOpen
  \bibfield  {author} {\bibinfo {author} {\bibfnamefont {C.~A.}\ \bibnamefont
  {Bertulani}}\ and\ \bibinfo {author} {\bibfnamefont {P.~G.}\ \bibnamefont
  {Hansen}},\ }\href {\doibase 10.1103/PhysRevC.70.034609} {\bibfield
  {journal} {\bibinfo  {journal} {Phys. Rev. C}\ }\textbf {\bibinfo {volume}
  {70}},\ \bibinfo {pages} {034609} (\bibinfo {year} {2004})}\BibitemShut
  {NoStop}%
\bibitem [{\citenamefont {Bertulani}\ and\ \citenamefont
  {Gade}(2006)}]{MOMDIS}%
  \BibitemOpen
  \bibfield  {author} {\bibinfo {author} {\bibfnamefont {C.}~\bibnamefont
  {Bertulani}}\ and\ \bibinfo {author} {\bibfnamefont {A.}~\bibnamefont
  {Gade}},\ }\href {https://doi.org/10.1016/j.cpc.2006.04.006} {\bibfield
  {journal} {\bibinfo  {journal} {Comp. Phys. Comm.}\ }\textbf {\bibinfo
  {volume} {175}},\ \bibinfo {pages} {372} (\bibinfo {year}
  {2006})}\BibitemShut {NoStop}%
\bibitem [{\citenamefont {Stroberg}\ \emph {et~al.}(2014)\citenamefont
  {Stroberg}, \citenamefont {Gade}, \citenamefont {Tostevin}, \citenamefont
  {Bader}, \citenamefont {Baugher}, \citenamefont {Bazin}, \citenamefont
  {Berryman}, \citenamefont {Brown}, \citenamefont {Campbell}, \citenamefont
  {Kemper}, \citenamefont {Langer}, \citenamefont {Lunderberg}, \citenamefont
  {Lemasson}, \citenamefont {Noji}, \citenamefont {Recchia}, \citenamefont
  {Walz}, \citenamefont {Weisshaar},\ and\ \citenamefont {Williams}}]{Stro14}%
  \BibitemOpen
  \bibfield  {author} {\bibinfo {author} {\bibfnamefont {S.~R.}\ \bibnamefont
  {Stroberg}}, \bibinfo {author} {\bibfnamefont {A.}~\bibnamefont {Gade}},
  \bibinfo {author} {\bibfnamefont {J.~A.}\ \bibnamefont {Tostevin}}, \bibinfo
  {author} {\bibfnamefont {V.~M.}\ \bibnamefont {Bader}}, \bibinfo {author}
  {\bibfnamefont {T.}~\bibnamefont {Baugher}}, \bibinfo {author} {\bibfnamefont
  {D.}~\bibnamefont {Bazin}}, \bibinfo {author} {\bibfnamefont {J.~S.}\
  \bibnamefont {Berryman}}, \bibinfo {author} {\bibfnamefont {B.~A.}\
  \bibnamefont {Brown}}, \bibinfo {author} {\bibfnamefont {C.~M.}\ \bibnamefont
  {Campbell}}, \bibinfo {author} {\bibfnamefont {K.~W.}\ \bibnamefont
  {Kemper}}, \bibinfo {author} {\bibfnamefont {C.}~\bibnamefont {Langer}},
  \bibinfo {author} {\bibfnamefont {E.}~\bibnamefont {Lunderberg}}, \bibinfo
  {author} {\bibfnamefont {A.}~\bibnamefont {Lemasson}}, \bibinfo {author}
  {\bibfnamefont {S.}~\bibnamefont {Noji}}, \bibinfo {author} {\bibfnamefont
  {F.}~\bibnamefont {Recchia}}, \bibinfo {author} {\bibfnamefont
  {C.}~\bibnamefont {Walz}}, \bibinfo {author} {\bibfnamefont {D.}~\bibnamefont
  {Weisshaar}}, \ and\ \bibinfo {author} {\bibfnamefont {S.~J.}\ \bibnamefont
  {Williams}},\ }\href {https://dx.doi.org/10.1103/PhysRevC.90.034301}
  {\bibfield  {journal} {\bibinfo  {journal} {Phys. Rev.C}\ }\textbf {\bibinfo
  {volume} {90}},\ \bibinfo {pages} {034301} (\bibinfo {year}
  {2014})}\BibitemShut {NoStop}%
\bibitem [{\citenamefont {Dickhoff}\ and\ \citenamefont
  {Barbieri}(2004)}]{Dick04}%
  \BibitemOpen
  \bibfield  {author} {\bibinfo {author} {\bibfnamefont {W.}~\bibnamefont
  {Dickhoff}}\ and\ \bibinfo {author} {\bibfnamefont {C.}~\bibnamefont
  {Barbieri}},\ }\href {\doibase https://doi.org/10.1016/j.ppnp.2004.02.038}
  {\bibfield  {journal} {\bibinfo  {journal} {Progress in Particle and Nuclear
  Physics}\ }\textbf {\bibinfo {volume} {52}},\ \bibinfo {pages} {377 }
  (\bibinfo {year} {2004})}\BibitemShut {NoStop}%
\bibitem [{\citenamefont {Duguet}\ \emph {et~al.}(2015)\citenamefont {Duguet},
  \citenamefont {Hergert}, \citenamefont {Holt},\ and\ \citenamefont
  {Som\`a}}]{Dugu15}%
  \BibitemOpen
  \bibfield  {author} {\bibinfo {author} {\bibfnamefont {T.}~\bibnamefont
  {Duguet}}, \bibinfo {author} {\bibfnamefont {H.}~\bibnamefont {Hergert}},
  \bibinfo {author} {\bibfnamefont {J.~D.}\ \bibnamefont {Holt}}, \ and\
  \bibinfo {author} {\bibfnamefont {V.}~\bibnamefont {Som\`a}},\ }\href
  {\doibase 10.1103/PhysRevC.92.034313} {\bibfield  {journal} {\bibinfo
  {journal} {Phys. Rev. C}\ }\textbf {\bibinfo {volume} {92}},\ \bibinfo
  {pages} {034313} (\bibinfo {year} {2015})}\BibitemShut {NoStop}%
\bibitem [{\citenamefont {Mutschler}(2015)}]{MutsPHD}%
  \BibitemOpen
  \bibfield  {author} {\bibinfo {author} {\bibfnamefont {A.}~\bibnamefont
  {Mutschler}},\ }\href {https://tel.archives-ouvertes.fr/tel-01206188}
  {\bibinfo {type} {{P}h.{D} thesis}},\ \bibinfo  {school} {{U}niversit\'e
  {P}aris {S}ud - {P}aris {XI}} (\bibinfo {year} {2015})\BibitemShut {NoStop}%
\bibitem [{\citenamefont {Spieker}\ \emph {et~al.}(2019)\citenamefont
  {Spieker}, \citenamefont {Gade}, \citenamefont {Weisshaar}, \citenamefont
  {Brown}, \citenamefont {Tostevin}, \citenamefont {Longfellow}, \citenamefont
  {Adrich}, \citenamefont {Bazin}, \citenamefont {Bentley}, \citenamefont
  {Brown}, \citenamefont {Campbell}, \citenamefont {Diget}, \citenamefont
  {Elman}, \citenamefont {Glasmacher}, \citenamefont {Hill}, \citenamefont
  {Pritychenko}, \citenamefont {Ratkiewicz},\ and\ \citenamefont
  {Rhodes}}]{spieker19}%
  \BibitemOpen
  \bibfield  {author} {\bibinfo {author} {\bibfnamefont {M.}~\bibnamefont
  {Spieker}}, \bibinfo {author} {\bibfnamefont {A.}~\bibnamefont {Gade}},
  \bibinfo {author} {\bibfnamefont {D.}~\bibnamefont {Weisshaar}}, \bibinfo
  {author} {\bibfnamefont {B.~A.}\ \bibnamefont {Brown}}, \bibinfo {author}
  {\bibfnamefont {J.~A.}\ \bibnamefont {Tostevin}}, \bibinfo {author}
  {\bibfnamefont {B.}~\bibnamefont {Longfellow}}, \bibinfo {author}
  {\bibfnamefont {P.}~\bibnamefont {Adrich}}, \bibinfo {author} {\bibfnamefont
  {D.}~\bibnamefont {Bazin}}, \bibinfo {author} {\bibfnamefont {M.~A.}\
  \bibnamefont {Bentley}}, \bibinfo {author} {\bibfnamefont {J.~R.}\
  \bibnamefont {Brown}}, \bibinfo {author} {\bibfnamefont {C.~M.}\ \bibnamefont
  {Campbell}}, \bibinfo {author} {\bibfnamefont {C.~A.}\ \bibnamefont {Diget}},
  \bibinfo {author} {\bibfnamefont {B.}~\bibnamefont {Elman}}, \bibinfo
  {author} {\bibfnamefont {T.}~\bibnamefont {Glasmacher}}, \bibinfo {author}
  {\bibfnamefont {M.}~\bibnamefont {Hill}}, \bibinfo {author} {\bibfnamefont
  {B.}~\bibnamefont {Pritychenko}}, \bibinfo {author} {\bibfnamefont
  {A.}~\bibnamefont {Ratkiewicz}}, \ and\ \bibinfo {author} {\bibfnamefont
  {D.}~\bibnamefont {Rhodes}},\ }\href {\doibase 10.1103/PhysRevC.99.051304}
  {\bibfield  {journal} {\bibinfo  {journal} {Phys. Rev. C}\ }\textbf {\bibinfo
  {volume} {99}},\ \bibinfo {pages} {051304} (\bibinfo {year}
  {2019})}\BibitemShut {NoStop}%
\bibitem [{\citenamefont {Caurier}\ \emph {et~al.}(2014)\citenamefont
  {Caurier}, \citenamefont {Nowacki},\ and\ \citenamefont {Poves}}]{SDPF-UMIX}%
  \BibitemOpen
  \bibfield  {author} {\bibinfo {author} {\bibfnamefont {E.}~\bibnamefont
  {Caurier}}, \bibinfo {author} {\bibfnamefont {F.}~\bibnamefont {Nowacki}}, \
  and\ \bibinfo {author} {\bibfnamefont {A.}~\bibnamefont {Poves}},\ }\href
  {\doibase 10.1103/PhysRevC.90.014302} {\bibfield  {journal} {\bibinfo
  {journal} {Phys. Rev. C}\ }\textbf {\bibinfo {volume} {90}},\ \bibinfo
  {pages} {014302} (\bibinfo {year} {2014})}\BibitemShut {NoStop}%
\end{thebibliography}%
\end{document}